\documentclass[acmsmall]{acmart}

\AtBeginDocument{%
  \providecommand\BibTeX{{%
    \normalfont B\kern-0.5em{\scshape i\kern-0.25em b}\kern-0.8em\TeX}}}

\setcopyright{acmcopyright}
\copyrightyear{2018}
\acmYear{2018}
\acmDOI{10.1145/1122445.1122456}

\usepackage{multirow}
\usepackage{xcolor}
\usepackage{fancyvrb}
\usepackage{paralist}
\usepackage{amsmath}
\usepackage{amssymb}
\usepackage{rotating}
\usepackage{graphics}
\usepackage{supertabular}
\usepackage{graphicx}
\usepackage{caption}
\usepackage{subcaption}
\PassOptionsToPackage{hyphens}{url}
\usepackage{hyperref}
\usepackage{comment}
\usepackage{ics_macros}
\usepackage[table]{colortbl}% http://ctan.org/pkg/xcolor

\usepackage{array}
\newcolumntype{L}[1]{>{\raggedright\let\newline\\\arraybackslash\hspace{0pt}}m{#1}}
\newcolumntype{C}[1]{>{\centering\let\newline\\\arraybackslash\hspace{0pt}}m{#1}}
\newcolumntype{R}[1]{>{\raggedleft\let\newline\\\arraybackslash\hspace{0pt}}m{#1}}

\newcommand{\specialcell}[2][c]{%
  \begin{tabular}[#1]{@{}c@{}}#2\end{tabular}}

\newtheorem{definition}{Definition}

\acmJournal{JACM}
\acmVolume{37}
\acmNumber{4}
\acmArticle{111}
\acmMonth{8}

\begin{document}

\title{Improving ICS Cyber Resilience through Optimal Diversification of Network Resources}


\author{Tingting Li}
\affiliation{%
 \institution{Imperial College London}
 \streetaddress{South Kensington}
 \city{London}
 \country{United Kingdom}}
\email{tingting.li@imperial.ac.uk}

\author{Cheng Feng}
\affiliation{%
  \institution{Siemens Corporate Technology}
  }
\email{cheng.feng@siemens.com}

\author{Chris Hankin}
\affiliation{%
 \institution{Imperial College London}
 \streetaddress{South Kensington}
 \city{London}
 \country{United Kingdom}}
\email{c.hankin@imperial.ac.uk}

%\renewcommand{\shortauthors}{Li, Feng and Hankin}

\begin{abstract}
Network diversity has been widely recognized as an effective defense strategy to mitigate the spread of malware. Optimally diversifying network resources can  improve the resilience of a network against malware propagation. This work proposes an efficient method to compute such an optimal deployment, in the context of upgrading a legacy Industrial Control System with modern IT infrastructure. Our approach can tolerate various constraints when searching for an optimal diversification, such as outdated products and strict configuration policies. We explicitly measure the \emph{vulnerability similarity} of products based on the CVE/NVD, to estimate the infection rate of malware between products. A \emph{Stuxnet}-inspired case demonstrates our optimal diversification in practice, particularly when constrained by various requirements. We then measure the improved resilience of the diversified network in terms of a well-defined \emph{diversity metric} and \emph{Mean-time-to-compromise (MTTC)}, to verify the effectiveness of our approach. We further evaluate three factors affecting the performance of the optimization, such as the network structure, the variety of products and constraints. Finally, we show the competitive scalability of our approach in finding optimal solutions within a couple of seconds to minutes for networks of large scales (up to 10,000 hosts) and high densities (up to 240,000 edges).
\end{abstract}




\maketitle

\section{Introduction}\label{sec:introduction}

Industrial Control Systems (ICS) are cyber-physical systems that are responsible for maintaining normal operation of industrial plants such as water treatment, gas pipelines, power plants and industrial manufacture. Modern industrial organizations perform an increasing large amount of operations across IT and Operational technology (OT) infrastructures, resulting in inter-connected ICS. It also creates new challenges for protecting such integrated industrial environments, and makes cyber-physical security threats even more difficult to mitigate \cite{nistguide}. Therefore, many industrial organizations started looking for methods to converge IT and OT infrastructures in more secure and resilient ways. In this paper, we consider software diversification as a way of deploying products across ICS and improving the resilience of the integrated systems. However, there are various real-world constraints we might encounter when finding an optimal diversification strategy, for instance, limited flexibility of diversification for legacy systems, strict configuration policies and other (un)desirable configuration requirements. Therefore, our approach particularly considers these constraints into optimization and evaluates the impact of these constraints on the optimal diversification.

Software mono-culture has been recognized as one of the key factors that promote and accelerate the spread of malware. It is widely acknowledged that diversifying network resources (e.g. software packages, hardware, protocols, connectivity etc.) significantly mitigates the infection of malware between similar products and reduces the likelihood of repeating application of single exploits \cite{hole2015diversity}. When facing attacks using zero-day exploits (i.e. unknown exploits), the situation becomes even worse as there are no available defense countermeasures to stop them. \emph{Stuxnet}, as the first cyber weapon against ICS, leveraged four zero-day vulnerabilities.  Until September 2010, there were about 100,000 hosts over 155 countries infected by Stuxnet \cite{stuxnet}. The invariability or high similarity of products used in most affected hosts accounts for the rapid infection and prevalence of Stuxnet. Therefore, diversity-inspired countermeasures have been introduced to improve the resilience of a network against malware propagation. However, it is not very clear about
\begin{inparaenum}[(i)]
\item how much diversification is required to reach an optimal/maximal resilience,
\item how exactly to deploy diverse resources across a network, and
\item how configuration constraints would harm the optimal diversification.
\end{inparaenum}
In this paper, we aim to mitigate stuxnet-like worm propagation by optimally diversifying resources. We consider a variety of off-the-shelf products to provide services at each host, and find the optimal assignment of them to maximize the network resilience.

There are two main trends of research investigating diversity as an effective defense mechanism. One trend seeks for solutions from software development such as n-version programming \cite{avizienis1985n}, program obfuscation \cite{bhatkar2003address} and code randomization \cite{pappas2012smashing}. The other trend studies diversity-inspired defense strategies from the perspective of security management. Specifically,  the goal of this trend is to find an optimal assignment of diverse products for each host in a network. For example, O'Donnell and Sethu proposed to diversify products in a network by a distributed coloring algorithm in \cite{o2004achieving}. Newell \emph{et al.} focused on diversifying routing nodes and found an efficient way to compute the optimal solutions in \cite{newell2015increasing}. A set of security metrics have been introduced by Zhang \emph{et al.} \cite{zhang2016network} to evaluate network diversity and its impact on the resilience against zero-day attacks. More details about related work are provided in Section \ref{sec:related}.

Our work lies in the second trend of research, in which we aim to find the optimal assignment of products to diversify a network. Most of the existing work has made three critical assumptions:
\begin{enumerate}[(i)]
\item there is \emph{no} configuration constraints when searching for an optimal assignment of products.
\item each node (or host) in a network was modelled by a \emph{single} label, indicating that there is only one vulnerable product (or service) running on a node, namely there is only one attack vector on each node that requires diversification.
\item each individual product shares \emph{no} vulnerability with any other, which implies that unique exploits are necessary to compromise different products.
\end{enumerate}
Nevertheless, we contend that these assumptions are unrealistic, and thus we drop these assumptions in this work. We specifically defined any configuration constraints into the process of optimization. Also we considered a more realistic infection model of malware. In the following subsection, a simple example demonstrates how these assumptions prevent us from modelling the actual infection model of malware.

In this paper, we start with formally defining the similarity of vulnerabilities between products to reflect the similar exploitability of products. We conduct a statistical study to estimate such vulnerability similarities by using public vulnerability databases such as Common Vulnerabilities and Exposures (CVE) \cite{cve} and the National Vulnerability Database (NVD) \cite{nvd}. Furthermore, we represent each host in a network by a multi-label node, which can be formally mapped to a discrete Markov Random Field (MRF) model. By combining the similarity metric and the MRF model,  we can construct the corresponding infection model of potential zero-day exploits across a network with a given product assignment. We then focus on computing an optimal product assignment to minimize the prevalence of zero-day exploits. Before our main contributions are enumerated in Section \ref{sec:contributions}, we present an illustrative example in Section \ref{sec:mot} to further explain the motivation.

\subsection{Motivational Example}
\label{sec:mot}
\begin{figure*}[!t]
  \centering
  \includegraphics[width=\textwidth]{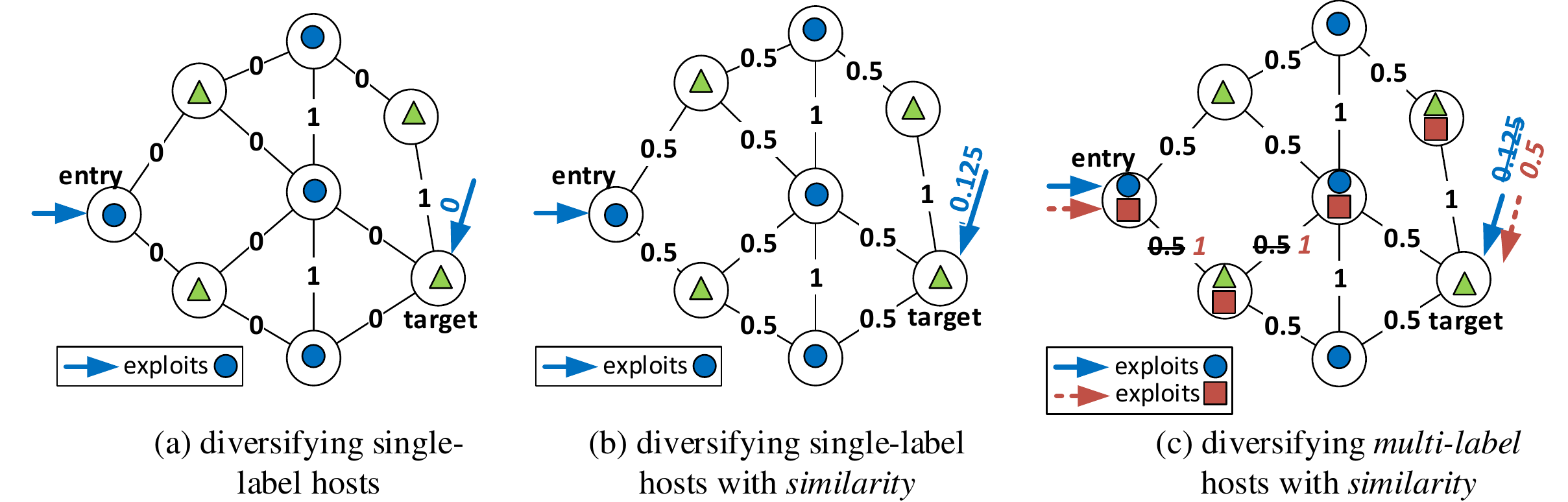}\\
  \caption{Motivational example about diversifying products in a network}\label{fig:mot}
\end{figure*}

We use a simple example in Figure \ref{fig:mot} to explain the motivation of this work, where a simplified network with 8 hosts is presented. Most of the existing work models the network as in Figure \ref{fig:mot}(a), where each host is represented by a single-label node. A zero-day exploit breaks into the network from the entry node. In order to prevent the exploit (which exploits circle labels) from infecting more hosts, most existing work suggests to diversify all hosts in the way indicated by triangle and circle labels respectively in Figure \ref{fig:mot}(a). The illustrated configuration is effective because the spread of the exploit is stopped after it compromises the entry node and hence the chance of the target node being infected is 0.

Nevertheless, it is assumed that different products share \emph{no} vulnerabilities between each other, which is however not always the case according to our statistical study on the CVE/NVD database in Section \ref{sec:sim}. We discover that most vulnerabilities reported to NVD can actually affect multiple products. Therefore, we improve the model by considering the vulnerability similarities between different products. Figure \ref{fig:mot}(b) demonstrates the zero-day propagation when the two products (circle and triangle labels) have a 0.5 vulnerability similarity between each other, namely there is a 50\% chance that the same zero-day vulnerability exploited at circle labels can also be exploited at triangle labels, and \emph{vice versa}. In this case, the probability of the target node being breached is increased to approximately 0.125.

In most realistic scenarios, a host is supposed to deliver multiple services (e.g. operating systems, web servers, email servers, databases, etc), each of which is potentially vulnerable to zero-day attacks. That means each host actually offers several alternative attack vectors, and as a result, sophisticated attackers can choose the vulnerability with higher success rate to exploit the host.  Therefore, we represent each host by multiple labels corresponding to different services on the host. As shown in Figure\ref{fig:mot}(c), we add another shape of labels (i.e. red squares) to some hosts, and introduce a sophisticated attacker in possession of \emph{two} zero-day exploits (one for round labels and the other for square labels). It can be seen from Figure\ref{fig:mot}(c), the attacker uses the square label exploit (rather than the round label exploit) to infect its adjacent node, which gives a greater chance of success. Consequently, with the collaboration of two zero-day exploits, the risk of the target node being compromised is further increased to approximately 0.5.

\subsection{Main Contributions} %\tlnote{minor changes of this section}
\label{sec:contributions}

From the example above, we learn that in order to find the optimal way to diversify network resources, we first need to model the resources accurately, based on which we can determine precisely the infection model of potential exploits across a network and find the optimal assignment of products to minimize the prevalence of exploits.  We summarize the main contributions of this paper as follow:

\begin{enumerate}[(i)]
  \item We demonstrate that our optimization approach is directly applicable in practice to find the optimal diversification strategy when integrating ICS with modern IT infrastructure. We use a real-world case study inspired by \emph{Stuxnet} propagation, to find optimal diversification solutions to IT-OT convergence of ICS, particularly accommodating real-world configuration constraints and limited flexibility of diversification in certain areas.
 \item To the best of our knowledge, our work is the first attempt to  explicitly consider the \emph{vulnerability similarity} between products when finding the optimal diversification solutions. Specifically, we represent each host by a \emph{multi-labelling model} with each label corresponding to a service on the host. A variety of products for each service is also modelled to render different assignments of products. By means of the vulnerability similarity between assigned products, the  infection of malware across the network can be accurately estimated.
 \item In order to compute the\emph{ optimal assignment} of products, we model the network by a \emph{discrete Markov Random Field} (MRF), which then can be optimized by an efficient \emph{sequential tree-reweighted message passing } (TRW-S) algorithm \cite{kolmogorov2015new}. In this way, our approach can scale up well to analyze large-scale and high-density networks.
\end{enumerate}

\subsection{Paper Organization}
%\tlnote{new subsection added, please review}
The rest of the paper is organized as follows. We discuss related work in Section \ref{sec:related}. The similarity metric is introduced in Section \ref{sec:sim}, as well as the statistical analysis based on CVE/NVD databases. We formally represent the network and the addressed research problem in Section \ref{sec:network}. The computation of optimal solutions is given in Section \ref{sec:opt}. A diversity metric based on Bayesian Networks (BN) is given in Section \ref{sec:metric} to evaluate our solutions. The case study about mitigating Stuxnet propagation in integrated ICS is presented in Section \ref{sec:stuxnet} to demonstrate the practical usage of our optimization approach, and an in-depth evaluation of our optimization approach is given in Section \ref{sec:eval}. The scalability analysis can be found in Section \ref{sec:random}. The paper finishes with a discussion and conclusion in Section \ref{sec:conclusion}.

\section{Related Work}\label{sec:related}
Software diversity has long been recognized as a mechanism for improving resilience and security of networked computing systems \cite{larsen2014sok,hole2015diversity,baudry2015multiple}. The rationale is that it forces attackers to develop an unique exploit to compromise an individual product at each node in a network, thus substantially increasing the attacking time and cost needed to penetrate a networked system at a massive scale.

A variety of methodologies for software diversity have been studied in the literature, among which the first direction of research focuses on software development diversity. Examples include n-version programming \cite{avizienis1985n}, execution environment diversity \cite{pal2014managed} and code randomization \cite{pappas2012smashing}.

The second direction, which is also the focus of this paper, is the strategies for diversified deployment of resources in a networked system. For instance, based on the assumption that different variants of products share no common vulnerabilities, O'Donnell and Sethu \cite{o2004achieving} proposed to assign diverse software packages in a communication network by a distributed coloring algorithm to limit the total number of nodes an attacker can compromise using a limited attack toolkit. Newell \emph{et al.}  \cite{newell2015increasing} found an efficient approach to compute the optimal solution for placing diverse software/OS variants on routing nodes to improve overall network resilience given the assumption that each variant is compromised independently with some probability metrics. Besides, there were some work defining formal security metrics for software diversity. For example, Wang \emph{et al.} \cite{wang2010k} defined a network security metric, k-zero day safety, for measuring the risk of unknown vulnerabilities based on the number of unknown vulnerabilities required for compromising network assets. Furthermore, Zhang \emph{et al.}  \cite{zhang2016network} introduced three security metrics to evaluate the resilience against zero-day attacks using different diversity strategies based on the number and distribution of distinct resources inside a network, the least attacking effort required for compromising certain important resources, and the average attacking effort required for compromising critical assets, respectively. Borbor \emph{et al.} \cite{borbor2016diversifying} explicitly considered cost constraints on optimizing software diversity strategies. It is noticed that most existing work assumes that there is a very limited attack surface provided at each host, namely there is only one vulnerable product at each host for attackers to exploit. By contrast, we explicitly model various attack vectors (offered by multiple products) at each host.% in this paper. , and attackers can choose any of them  to exploit.

Vulnerability databases such as CVE/NVD can provide statistical evidence for measuring software diversity. For example, Garcia \emph{et al.} \cite{garcia2011diversity} presented a study on the overlap of vulnerabilities in 11 different OSes with OS vulnerability data from NVD. In \cite{bozorgi2010beyond}, Bozorgi \emph{et al.} trained classifiers to predict whether and how soon a vulnerability is likely to be exploited by applying machine learning techniques on CVE data. On the validity issue of CVE/NVD, Johnson \emph{et al.} conducted the  assessment of several well-known vulnerability databases and concluded that NVD was actually the most trustworthy database \cite{johnson2016can}; we used NVD in this paper.

%\tlnote{this paragraph needs mention again in model section}
Some existing work \cite{wang2010k}\cite{zhang2016network}\cite{li2016critis} studied malware propagation based on attack graphs to assess the risk of malware along with specific attack paths and network topology. Attack graphs have been extensively studied in the community to express the exploitation conditions of vulnerabilities. However, due to the unknown nature of zero-day vulnerabilities, we contend that such approaches are not always feasible to model zero-day malware. In contrast to existing work using attack graphs, our work focuses on the speed of zero-day exploits across a network configured by similar products. Highly similar configurations (in terms of potential vulnerabilities) would accelerate the prevalence of zero-day exploits. Instead of producing specific propagation paths, we use more general undirected edges to symbolize the connections (rather than directed information flow) between different hosts. We then use the proposed similarity metric to estimate the infection of zero-day exploits on each edge and find optimal diversification solutions.% As a result, we can construct the infection model of zero-day exploits in order to estimate to what degree a zero-day exploit would compromise a network by infecting hosts as extensively/quickly as possible.

\section{Similarity of Product Vulnerability} \label{sec:sim}
%\tlnote{no changes made in this section}

In this section, we formally introduce the notion of vulnerability similarity between a pair of products, namely the likelihood of an exploit compromising both products.

\begin{definition}[Similarity of Product Vulnerability]\label{def:sim-host}
Let $x_i$, $x_j$ be a pair of products, $\bold{V}_{x_1}$  and  $\bold{V}_{x_j}$ are sets of vulnerabilities of  $x_i$ and $x_j$ respectively. The vulnerability similarity between $x_i$ and $x_j$ can be obtained by the \emph{Jaccard similarity coefficient \cite{choi2010survey}}: %$ sim(x_i, x_j) =  \frac{| \bold{V}_{x_i} \cap\bold{V}_{x_j} | }{| \bold{V}_{x_i} \cup \bold{V}_{x_j}|}$
\begin{equation*}\label{equ:sim}
   sim(x_i, x_j) =  \frac{| \bold{V}_{x_i} \cap\bold{V}_{x_j} | }{| \bold{V}_{x_i} \cup \bold{V}_{x_j}|}
\end{equation*}
\end{definition}
Given a pair of products, the vulnerability similarity is estimated by the ratio of the number of shared vulnerabilities between the two products to the total number of vulnerabilities. The rationale for this is to capture statistically how similar the vulnerabilities found on two products are. %\tlnote{needs better wording}

To provide a more realistic sense, we can use the data from the NVD database \cite{nvd} to calculate the similarity metric for any pair of off-the-shelf products. There are more than 116,120 vulnerabilities published by NVD at the time of this paper. Each NVD vulnerability feed contains information about a specific vulnerability. An example of an NVD entry is given in Table \ref{tab:cve}, which includes a unique identifier by the Common Vulnerability Enumeration (CVE) with the format ``\texttt{CVE-YEAR-NUMBER}'', the attack scenarios using the vulnerability, and the affected products sorted by Common Platform Enumerations (CPEs). CPE provides a well-formed naming scheme for IT systems, platforms and packages. %Each CPE follows a general format ``\texttt{cpe:/PART:VENDOR:PRODUCT:VERSION}''. The \texttt{PART} specifies if the CPE refers to hardware (\texttt{h}), an operating system (\texttt{o}) or an application (\texttt{a}).
Table \ref{tab:cve} shows that if the vulnerability \texttt{CVE-2016-7153} is exploited, a number of web browsers from different vendors are affected such as \emph{IE, Chrome, Firefox, Opera, etc}. %\tlnote{restored the explanation of table 1}

\begin{table}[!h]
{\scriptsize
\centering
\caption{Simplified NVD Vulnerability Summary for \texttt{CVE-2016-7153}}
\label{tab:cve}

\begin{tabular}{|p{1.4cm}|p{10.4cm}|}
\hline
\textbf{CVE-ID }                          & CVE-2016-7153                                                                                                                                                                                                                                                                                                                                                                                                                                                                      \\ \hline
\textbf{Overview }                        & The HTTP\/2 protocol does not consider the role of the TCP congestion window in providing information about content length, which makes it easier for remote attackers to obtain cleartext data by leveraging a web-browser configuration in which third-party cookies are sent, aka a "HEIST" attack. \\ \hline
\textbf{Release Date} & September 6th, 2016 \\ \hline
\textbf{CVSS v3.0 } \newline \textbf{Severity and Metrics:}                    & Base Score: \emph{5.3 MEDIUM }; \hspace{5mm} Vector: \emph{AV:N/AC:L/PR:N/UI:N/S:U/C:L/I:N/A:N} ; \newline
                                   Impact Score: \emph{1.4 }; \hspace{1.25cm} Exploitability Score: \emph{3.9}   \newline                                                                                                                                                                                                                                                                                                                                                          Access Vector: \emph{Network};  \hspace{7mm} Access Complexity: \emph{Low }; \hspace{5mm}Privileges Required (PR):\emph{ None}; \newline
                                   User Interaction (UI): \emph{None}; \hspace{3.5mm} Scope (S): \emph{Unchanged};  \hspace{7.5mm} Confidentiality (C):\emph{ Low}; \newline
                                   Integrity (I): \emph{None};\hspace{1.35cm}  Availability (A): \emph{None} 
                                   \\ \hline
\textbf{Vulnerable software} \newline \&\textbf{Versions} &
\begin{tabular}[c]{l}
\texttt{cpe:/a:microsoft:edge:-} \\ \texttt{cpe:/a:microsoft:internet\_explorer:-} \\
\texttt{cpe:/a:google:chrome:-} \\ \texttt{cpe:/a:apple:safari} \\
\texttt{cpe:/a:mozilla:firefox} \\ \texttt{cpe:/a:opera:opera\_browser:-}
\end{tabular}                                                                                                                                              \\ \hline
\end{tabular}}
\end{table}

%\tlnote{restored the original table 1 }

Given the large number of vulnerabilities in NVD, CPE  serves the role of sorting vulnerabilities according to their affected products. We developed a program based on \textsc{cve-search}\cite{cve-search} to fetch necessary data from NVD, filter out vulnerabilities for each studied product, and calculate the similarity of vulnerabilities between products. The  pairwise similarities are stored as \emph{Similarity Tables}. In this way, we can calculate the similarity of vulnerabilities between \emph{any} pair of products listed in NVD.

For the purpose of illustration, here we use operating systems and web browsers as examples. We collect vulnerabilities for 9 common OS products and 8 common web browsers reported in the period between 1999 and 2016. Table \ref{tab:os-vuln} enumerates the pairwise similarity between the chosen OS products and Table \ref{tab:wb-vuln} for the chosen web browsers. The reason for choosing these products is mainly because they have been ranked as most vulnerable products by \emph{CVE Details} \cite{cve-details1}. Each entry of the two tables contains the pairwise similarity  calculated by Def.(\ref{def:sim-host}) and the number of shared vulnerabilities between products in brackets. The diagonal entries in tables are the total number of vulnerabilities of the row/column product. As the pairwise similarity is symmetric, the other half of a similarity table is omitted. For reserving the generality and flexibility of our study, we implicitly consider each different release/version of a product as a distinct \emph{product} to compare. For instance, \emph{Windows 8.1} and \emph{Windows 7} are treated as two individual products and sorted by two different CPE queries \texttt{cpe:/o:microsoft:windows\_7} and \texttt{cpe:/o:microsoft:windows\_8.1}.

%\tlnote{emphasis the reason of choosing old OS.}

From the tables, we can observe that products of the same vendor tend to have higher similarities. Two exceptions are observed in Table \ref{tab:os-vuln}: \emph{Mac OS X 10.5} and \emph{Windows 7} shares 8.1\% vulnerabilities, and \emph{Ubuntu 14.04} and \emph{Debian 8.0} have 20.8\% vulnerabilities in common, despite these two pairs of products being from different vendors. It is also noticed that products with a longer gap between their release dates have a lower similarity. 

Based on the statistical studies in both tables, we conclude that a single vulnerability is likely to affect multiple products across different versions, different vendors and different platforms, which implies that a single zero-day vulnerability could probably be exploited on heterogeneous hosts in a network. Therefore, to maximize the resilience of a network against zero-day exploits, it is desirable to use the \emph{up-to-date} products from \emph{diverse} vendors across a network. For instance, \emph{Windows 10} has much lower similarities of vulnerabilities with the other Windows OS, and even shares no vulnerability with \emph{Windows XP}. However, it is not always feasible to deploy the latest and diverse products due to their incompatibilities with other services. For instance, SIMATIC WinCC is one of the most widely applied SCADA systems, but it can only operate on Windows OS, and most releases of WinCC do not fully support \emph{Windows 10} yet \cite{wincc}. %\tlnote{incorrect! check WinCC manual}.

It is worth mentioning that the versions of selected software in both tables are constrained by the granularity of CPE search engine. The CPE entries for many vulnerabilities in NVD are not complete or of different granularities. %, i.e. not all the vulnerable versions for a CVE are listed.

In this section, we demonstrated the usage of CVE data to calculate the vulnerability similarity. The NVD database is one of the most well-known publicly accessible vulnerability databases, which also covers most off-the-shelf products and up-to-date vulnerability information. %However, it should be noticed that the NVD database is not the only source to calculate the vulnerability similarity. %In our ongoing research work, we are always looking for alternative ways to measure the similarity.

%\tlnote{reduced up to here...}
%\tlnote{if space allowed, will add a new table with average \# affected vendors, \#CPEs, \#affected products}

\begin{table*}[!t]
\centering
{\footnotesize
\centering
\caption{Similarity Table for Common OS Products from CVE/NVD }
\label{tab:os-vuln}
\scalebox{0.95}{
\begin{tabular}{c||c|c|c|c|c|c|c|c|c}
\hline
 & WinXP2  &   Win7    & Win 8.1   & Win10      & Ubt14.04 & Deb8.0  & Mac10.5 & Suse13.2 & Fedora \\ \hline \hline
WinXP2                     & 1.00 (479) &            &            &             &            &            &             &         &   \\ \hline
%\textbf{WinXP SP3}         & 0.57 \emph{(448)} & 1.00 (752) &            &             &            &            &             &         &   \\ \hline
%Win2003                                 & 0.59 (395) &  1.00 (580) &             &            &            &             &            &\\ \hline
Win7                                    & 0.278 (328) &  1.00 (1028) &            &            &             &            &       &      &\\ \hline
Win8.1                                  & 0.009 (10)  &   0.228 (298)  & 1.00 (572) &            &             &            &        &      &\\ \hline
Win10                                   & 0 (0)      &  0.124 (164)  & 0.697 (421) & 1.00 (453) &             &            &        &      &\\ \hline
Ubt14.04                             & 0 (0)      &  0 (0)       & 0 (0)      & 0 (0)      & 1.00 (612)  &            &        &      &\\ \hline
Deb8.0                               & 0 (0)      & 0 (0)       & 0 (0)      & 0 (0)      & 0.208(195)   & 1.00 (519) &        &      &\\ \hline
Mac10.5                               & 0 (0)      & 0.081 (109)       & 0 (0)      & 0 (0)      & 0 (0)   & 0 (0) &  1.00(424)      &      &\\ \hline
Suse13.2                               & 0 (0)      & 0 (0)       & 0 (0)      & 0 (0)      & 0.170(161)   & 0.112 (102) & 0 (0)       & 1.00(492)     &\\ \hline
Fedora                              & 0 (0)      & 0 (0)       & 0 (0)      & 0 (0)      & 0.083(75)   & 0.049 (41) &  0.001(1)      &  0.116 (89)    & 1.00(367) \\ \hline
\end{tabular}}}
\end{table*} 
\begin{table*}[!h]
\centering
\caption{Similarity Table for Common Web Browser from CVE/NVD}
\label{tab:wb-vuln}
\scalebox{0.85}{
\begin{tabular}{c||c|c|c|c|c|c|c|c}
\hline
\multicolumn{1}{c|}{} & \multicolumn{1}{c|}{IE8} &  \multicolumn{1}{c|}{IE10} & \multicolumn{1}{c|}{Edge} & \multicolumn{1}{c|}{Chrome} & \multicolumn{1}{c|}{Firefox} & \multicolumn{1}{c|}{Safari} & \multicolumn{1}{c|}{SM} & \multicolumn{1}{c}{Opera} \\ \hline \hline
IE8                    & 1.0 (349)               &                           &                          &                               &                                &                &              &\\ \hline
%IE9                    & 0.52 (282)               & 1.00 (471)               &                           &                           &                           &                               &                                &                              \\ \hline
IE10                   & 0.386 (240)               & 1.0 (513)                &                           &                               &                                &                &              &\\ \hline
%IE11                   & 0.28 (188)               & 0.45 (305)               & 0.65 (405)                & 1.00 (513)                &                           &                               &                                &                              \\ \hline
Edge                   & 0.014 (7)                 & 0.121 (73)                 & 1.0 (194)                 &                               &                                &                 &             &\\ \hline
Chrome               & 0 (0)                    & 0 (0)                     & 0.001 (2)                     & 1.0 (1661)                     &                                &                 &         &    \\ \hline
Firefox              & 0 (0)                    & 0 (0)                     & 0.001 (2)                     & 0.005 (15)                         & 1.0 (1502)                      &                 &            & \\ \hline
Safari                & 0 (0)                   & 0 (0)                     & 0.002 (2)                     & 0.009 (21)                         & 0.003 (6)                          & 1.0 (766)         &        &   \\ \hline
SeaMonkey           & 0 (0)              & 0 (0)                     & 0 (0)                    & 0.001 (3)                         & 0.450 (683)                          &    0.001(1)      &    1.0(492)    &  \\ \hline
Opera              & 0 (0)              & 0 (0)                     & 0.003 (1)                    & 0.003 (6)                         & 0.004 (7)                          & 0.004(4)         &    1.00(492) & 1.00(225)      \\ \hline
\end{tabular}}
\end{table*}

\section{Diverse Product Assignment}\label{sec:network}
%\tlnote{no changes made in this section}
In this section, we present the formal model of a product assignment for a given network, which is to find a diversification solution to assigning products to each host such that the malware propagation between similar products can be effectively mitigated. % First of all, we introduce the network model used in this paper, as well as the product assignment and configuration constraints. %\tlnote{not yet completed}

Each host has to provide a set of services $S$, such as an operating system, a web browser and a database server. Each service can be provided  by a range of diverse products $P$. Therefore, we formally define a network in terms of hosts, links, services and products as below.

\begin{definition}[\textbf{Network}]
Let $N = \langle H, L, S, P \rangle$ be a network,
\begin{itemize}
  \item $H$ = $\{\host{0}, \dots, \host{n} \} $  is a set of hosts.
  \item $L$ captures the links between a pair of hosts, $L \subseteq H \times H $
  \item $S$ = $\{s_{1}, \dots, s_{m} \} $  is a set of services, and $\sh{i}\in 2^S$ denotes a set of services provided by a host $\host{i}$.
  \begin{equation}\label{def:products}
   \sh{i} = \{ \s{1}, \dots, \s{k} \} , ~~\text{where}~~ \sh{i} \in 2^S, k \leqslant  m
  \end{equation}
  \item $P$ is a set of off-the-shelf products, and hence each service $\s{j}$ can be provided by a range of diverse products,
  \begin{equation}\label{def:service}
    p(\s{j}) = \{ \p{j}{1}, \dots, \p{j}{l} \},~~\text{where}~~ \p{j}{x}\in P.
  \end{equation}
\end{itemize}
\end{definition}

Assigning one product for each service on a host is termed as \emph{an assignment of products} for a host. %The formal definition is given as below.

\begin{definition}[\textbf{Product Assignment}]
Given a network $N = \langle H, L, S, P \rangle$, an assignment of products is captured by $\as': H \times S \rightarrow P$, such that $\as'(\host{i}, \s{j})$ is the product assignment for a service $\s{j} \in \sh{i}$ at the host $\host{i}$: $\as'(\host{i}, \s{j})  = \p{j}{x}$. The assignment for all services at a host $\host{i} \in H$ can be derived by $\as: H \times 2^S \rightarrow 2^P$:
\begin{eqnarray*}
\as(\host{i}, \sh{i})& = & (\as'(\host{i}, \s{1}),\dots \ \as'(\host{i}, \s{k})) \\
   & = &  ( \p{1}{m}, \dots, \p{k}{n} ) \\
  && \hspace{-18mm} \text{where}~~\p{1}{m} \in p(s_1),\dots,~ \p{k}{n} \in p(s_k)
\end{eqnarray*}
\end{definition}

\begin{figure}[!t]
  \centering
  \includegraphics[width=0.6\textwidth]{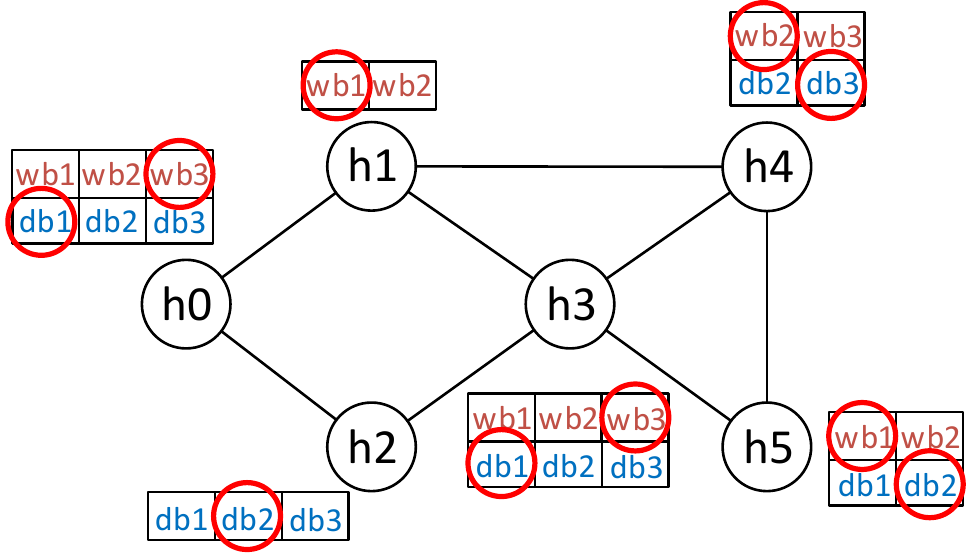}\\
  \caption{A network with an assignment $\alpha$ by red circles}\label{fig:toy}
\end{figure}

Therefore $\as$ allocates products to all services running on a host, whilst $\as'$ assigns a product to a specific service of a host. An example network is illustrated in Figure \ref{fig:toy}, where a network consisting of 6 hosts $H = \{\host{0},\dots \host{5}\}$ is modelled. Each host provides up to two essential services  \emph{web browser} and \emph{database}. Three diverse web browser products \{$wb_1, wb_2, wb_3$\} and three database products \{$db_1, db_2, db_3$\} are available to choose. Each host might have different ranges of products to choose. A \emph{possible} product assignment $\alpha$ is highlighted by red circles in Figure \ref{fig:toy}

Now the problem is to find an optimal assignment which allocates \emph{most diverse} products for each pair of connected hosts, so that the likelihood of a malware propagation between two hosts can be minimized. Nevertheless, some configuration requirements might hinder us from choosing the most optimal product assignment in practice. Therefore, we formally introduce \emph{local} and \emph{global} constraints to represent those requirements into the optimization process.   %a particular host is required to run Linux operating systems only due to any.

A \emph{local constraint} indicates that for a particular host, a product $p_j$ is required to either configure with another product $p_l$ (expressed by $c_y$), or avoid the product $p_k$ (expressed by $c_x$). Such requirements can also be applied to all hosts by using \emph{global constraints}.

\begin{definition}[\textbf{Configuration Constraints}]\label{def:constraints}
Given a network $N = \langle H, L, S, P \rangle$, a set of constraints $\mathcal{C}$ expresses any (un)desirable product combinations in the solution. A constrained solution $\as_{\mathcal{C}}$ allocates products subject to $\mathcal{C}$.
\begin{itemize}
\item a \emph{local constraint} is applied to a specific host $\host{i} \in H$ in the form of: $c_x := \langle\host{i}, \s{m}, \s{n}, +p_j, -p_k\rangle$  or  $c_y := \langle\host{i}, \s{m}, \s{n}, +p_j, +p_l\rangle$  such that the constrained solution $\as_{\mathcal{C}}$ satisfies:
    \begin{eqnarray*}
         \forall c_x \in \mathcal{C}&:& \as'_{\mathcal{C}}(h_i, s_{m} ) = p_j  ~\land~  \as'_{\mathcal{C}}(h_i, s_{n}) \neq p_k  \\
         \forall c_y \in \mathcal{C}&:& \as'_{\mathcal{C}}(h_i, s_{m} ) = p_j  ~\land~  \as'_{\mathcal{C}}(h_i, s_{n}) = p_l
    \end{eqnarray*}
\item a \emph{global constraint} is applied to all hosts in $H$ in the form of: $c_x := \langle\textsf{ALL}, \s{m}, \s{n}, +p_j, -p_k\rangle$  or  $c_y := \langle\textsf{ALL}, \s{m}, \s{n}, +p_j, +p_l\rangle$  such that $\as_{\mathcal{C}}$ satisfies:
    \begin{eqnarray*}
         \forall c_x \in \mathcal{C}\;, \forall h_i \in H: \as'_{\mathcal{C}}(h_i,  s_{m} )= p_j  \land \as'_{\mathcal{C}}(h_i, s_{n}) \neq p_k \\
         \forall c_y \in \mathcal{C}\;, \forall h_i \in H: \as'_{\mathcal{C}}(h_i,  s_{m} )= p_j  \land\as'_{\mathcal{C}}(h_i, s_{n} )= p_l
    \end{eqnarray*}
\end{itemize}
\end{definition}

%For instance, we can specify a global constraint $c_1 = \langle\textsf{ALL}, \s{os}, \s{wb}, + wb_3, - db_1 \rangle$ to avoid . To comply with  $c_1$, the illustrated assignment $\as$ on $h_0$ and $h_3$ in Figure \ref{fig:toy} would be updated.

The usage of constraints is demonstrated in the later case study (Section \ref{sec:opt-solution}). Now we can define the optimal assignment of products $\opm{\as}$ and the constrained optimal assignment $\opm{\as}_\mathcal{C}$ as follows. %In this next section, we continue to discuss how to find those optimal assignments.

\begin{definition}[\textbf{Optimal Diversification}]
Given a network $N = \langle H, L, S, P \rangle$, an \emph{optimal} assignment of products is captured by $\opm{\as}: H \times 2^S \rightarrow 2^P$, such that $\opm{\as}(\host{i}, \sh{i})$ is the optimal product assignment for a host $\host{i} \in H$. A \emph{constrained optimal} solution is denoted by  $\opm{\as}_\mathcal{C}$ which provides an optimal product assignment subject to a set of local and global constraints $\mathcal{C}$.
\end{definition}

We adopt the following notation convention throughout this paper. $\as$ denotes an assignment of products for a network in general. $\opm{\as}$ is for an optimal
assignment \emph{without} constraints, and $\opm{\as}_\mathcal{C}$ is for a constrained optimal assignment. Specifically, $\as(\host{i}, \sh{i})$ includes the products
assigned to a host $h_i$, and $\as'(\host{i}, \s{m})$ is the product assigned to a particular service $s_m$ at the host $h_i$.

In the next section, we focus on finding such an optimal assignment of products $\opm{\as}$ for a given network, as well as computing constrained optimal solutions in Section \ref{sec:cons}.

\section{Finding the Optimal Diversification}\label{sec:opt}

%\tlnote{this paragraph is added to connect MRF with the preceding section}
First of all, we need a model  to accurately represent a network in which each host has multiple services and each service can be provided by a range of products. More importantly,  this model has to offer sufficient flexibility, because each host runs a customized set of services and even the same service has various selections of products at different hosts due to any compatibility requirements. Furthermore, we have to consider whether there is any existing efficient optimization algorithm to such a model. For these purposes, the optimal diversification problem can be represented by using a discrete Markov Random Field (MRF), which is converted into an optimal assignment problem of MRF that can be solved by an efficient message passing algorithm.
%Consequently, we choose to formally model the problem by using discrete Markov Random Field (MRF), which  perfectly fits all our requirements.  The optimal diversification problem is then converted  into an optimal assignment problem of MRF that can be solved by an efficient message passing algorithm. %given the notations from the previous section, there is a set of services from $S$ for each host $\host{i} \in H$ in our network, and we need to determine the product labels assigned to each service in an optimal way.

Specifically, we model this problem as a discrete MRF where each host has up to  $|S|$ services, and there are up to $|P|$ products for each service $\s{k} \in S$. The optimization assigns up to $|S|$ products -- one for each service on each host --  to reach the global minima of the propagation.
Given a network $N = \langle H, L, S, P \rangle$, we derive the energy function $\bold{E}$ to denote the \emph{unary cost} for each host and \emph{pairwise cost} between a pair of connected hosts.
\begin{equation}\label{eq:crf}
\bold{E}(N)= \sum_{\substack{\host{i} \in H \\\s{k}\in S_{\host{i}}}} \phi (\host{i},\s{k})+\sum_{(\host{i},\host{j})\in L}\psi  (\as(\host{i}, S_{\host{i}}),  \as(\host{j}, S_{\host{j}}) )
\end{equation}
\noindent where $\phi(\cdot)$ denotes how likely a product is preferred by a host $\host{i}$ to deliver the service $\s{k}$, and $\psi(\cdot,\cdot)$ is a pairwise cost between the products assigned to a pair of connected hosts, which in our context would be the pairwise similarity between products. Our problem is then mapped to the context of \emph{Conditional Random Fields}~\cite{geman1984stochastic}, with regard to a minimum of energy $\bold{E}$ corresponding to a \emph{maximum a-posteriori} (MAP) labeling of the service $\s{k}$. In the following subsections, we discuss the formulation of the unary cost $\phi(\cdot,\cdot)$ and pairwise cost $\psi(\cdot,\cdot)$ in more detail.

\subsection{Unary Cost $\phi(\cdot)$}

The unary cost is derived from the preference of a specific product for a host. %In general cases, such preference should be penalized with a lower probability as they are more likely to be targeted by attackers.
By considering one product being assigned to each host, our unary cost  $\phi(\cdot)$ is expressed as
\begin{equation}
 \sum_{\host{i} \in H}  \sum_{\s{k} \in \sh{i}} Pr(\as'(\host{i}, \s{k}) | \host{i} )
\end{equation}
\noindent where $Pr(\cdot)$ presents the probability that a product is assigned to $\host{i}$. In general cases, there is no specific preference amongst available products for each host to deliver a service. Therefore this term   can be replaced by a small constant $Pr_{const}$ for optimization. Although such a unified cost provides fast convergence in optimization, the real-world networks would be more complex and constrained by practical requirements as discussed in Section \ref{sec:network}. Therefore, the unary cost is further refined subject to any constraints in the next subsection.

\subsection{Unary Cost $\phi(\cdot)$ with Constraints}\label{sec:cons}

In our system, constraints are implemented as conditional patches to our energy function, in particular the unary cost. For a local constraint $c \in \mathcal{C}$ expressing an undesirable requirement  $c:= \langle h_i, \s{m}, \s{n}, +p_j, -p_k\rangle$  or a desirable requirement $c:= \langle h_i, \s{m}, \s{n}, +p_j, +p_k\rangle$, our unary cost $\phi(\cdot)$ can be represented as follows:
\begin{eqnarray*}
Pr(\as'(\host{i}, \s{j}) | \host{i} ) \hspace{5.8cm}\\ =
\begin{cases}
P_c(\alpha|\alpha'(h_i,s_m)=p_j,\alpha'(h_i,s_n)=p_k) & \text{if}~ \s{j} = \s{m} \\
Pr_{const} & \scriptsize otherwise
\end{cases}
\end{eqnarray*}

For unconstrained services, there is no preference amongst products and the unary cost is given by  a small constant $Pr_{const}$. For the constrained services (when $\s{j} = \s{m}$ ), the unary cost is given by $P_c(\cdot)$ which is interpreted as below according to the two types of local constraints above:
%where we keep the assumption that no preference on unconstrained hosts by using a small constant as the cost of assignment. F given two types of local constraints above, we have:
\begin{eqnarray*}
P_c(\alpha|\alpha'(h_i,s_m)=p_j,\alpha'(h_i,s_n)=p_k)  \\
\propto\begin{cases}
0 & \text{if}~~c:= \langle h_i, \s{m}, \s{n}, +p_j, +p_k\rangle\\
\infty & \text{if}~~c:= \langle h_i, \s{m}, \s{n}, +p_j, -p_k\rangle
\end{cases}
\end{eqnarray*}
\noindent where the desirable constraint contributes no additional cost whilst the undesirable constraint introduces a large cost. In this case, the optimization is induced to reach desirable assignments, but avoid undesirable ones in energy minimization. Note that such customized unary cost can also be applied for any global constraint, which is equivalent to applying a local constraint to all hosts.

\subsection{Pairwise Cost $\psi(\cdot,\cdot)$}

The pairwise  cost is derived from the similarity between the assigned products that provide the same service. As mentioned previously, a pair of connected hosts being assigned with more similar products would have greater infection rate, namely a zero-day exploit at one host is more likely to infect the other. When defining the pairwise cost, we penalize such similarities in order to provide a more diverse product assignment for the network. To achieve that, we define the pairwise cost term $\psi(\cdot,\cdot)$ as:

\begin{equation}
\sum_{(\host{i},\host{j})\in L}\sum_{\s{k} \in \sh{i} \cap \sh{j}} sim(\as'(\host{i}, \s{k}), \as'(\host{j}, \s{k}))
\end{equation}

\noindent where $\host{i}$ and $\host{j}$ denote a pair of connected hosts, and $sim(\cdot,\cdot)$ presents the similarity between two products providing the same service on a pair of connected hosts. It serves as a strong regularization on the product assignment as it ideally prevents the same product from being assigned to connected hosts.

\subsection{Energy Optimization}\label{sec:energy-opt}

Based on the unary cost and pairwise cost, we can determine the optimal assignment $\opm{\as}$ for $N$ by minimizing the energy function as below:
\begin{eqnarray*}
  \opm{\as} &=&  \operatorname*{argmin}_{\as} \bold{E}(N) \\
    &=& \operatorname*{argmin}_{\as} \sum_{\host{i} \in H}  \sum_{\s{j} \in \sh{i}} Pr(\as'(\host{i}, \s{j}) | \host{i} ) \\
    && +\sum_{(\host{i},\host{j})\in L}\sum_{\s{k} \in \sh{i} \cap \sh{j}} sim(\as'(\host{i}, \s{k}), \as'(\host{j}, \s{k}))
\end{eqnarray*}
\normalsize

%In our scenario, we propagate the pairwise similarity between two products (Sec 3.3).

%\noindent where $\widehat{\graph} = (H, L, \widehat{\as})$ is the optimal epidemic model subject to an optimal product assignment $\widehat{\as}$.
\noindent Solving such an energy is NP-Hard, and the alternative way is to use an approximate minimization algorithm to achieve a solution.  The well-known techniques for solving such problems are based on \emph{graph-cuts} and \emph{belief propagation (BP)}. The former is currently considered as the most accurate minimization approach for energy functions arising in many complex scenarios but it can be applied to a limited range of energy forms. If the form  is outside the class, like our energy function in Eq.~\ref{eq:crf}, BP is the common alternative. However, BP might not converge when applying to a wide range of convex functions. Instead, we employ a  \emph{sequential tree-reweighted message passing algorithm (TRW-S)}~\cite{kolmogorov2015new}. Similar to BP, TRW-S can be applied to the type of problems with the energy form in Eq.~\ref{eq:crf}. It is also guaranteed to give an \emph{optimal} MAP solution in most cases~\cite{kolmogorov2015new}. TRW-S outperforms BP and graph-cuts on many heavy tasks. It also demonstrates a great potential for the cases of labeling of nearly flat probabilities, as well as the cases of large-scale networks.

Our optimization scheme mainly follows \cite{kolmogorov2015new}, which is also extended to a multi-level fashion to better fit our problem.
Specifically, we enable the possibility of the parallel computation and even GPU acceleration. In addition, the optimization of the constrained
energy is straightforward because our constraints are efficiently encoded into the unary cost by manipulating the cost for specific hosts and assignments.
More details about the scalability analysis are given in Section \ref{sec:random}. A case study using our optimization approach in practice can be found in the later
Section \ref{sec:stuxnet}.

\section{Evaluation of Network Diversity}\label{sec:metric}
%\tlnote{no changes made in this section}
%Having presented our approach to finding an optimal diversification solution in the last section, this section focuses on using a well-studied network diversity metric to \emph{evaluate} our optimization solution.

The purpose of this section is to evaluate how much diversity a specific product assignment can bring about into a network, and we achieve this by using a network diversity metric based on Bayesian Networks \cite{zhang2016network}. Given a network $N$ and a specific product assignment $\as$,  we first construct its corresponding Bayesian Network (in Section \ref{sec:bn}) to estimate the infection rate on each edge between hosts, based on which we can evaluate the network diversify by calculating the value of the metric (in Section \ref{sec:div}).

%The metric was originally defined based on a Bayesian Network (Section \ref{sec:bn}), and we further adapt the metric to be applicable in our model for evaluation in Section \ref{sec:div}.

\subsection{Bayesian Network Evaluation Model}\label{sec:bn}
Before we define the complete Bayesian Network, we need a way to capture the impact of the attacker's behaviour on malware propagation. From an attack entry host, there are different ways to reach the final target by continuously exploiting a number of stepping-stone hosts. At each attack step from one host to another, there are often more than one vulnerable products to exploit and induce further spread of the malware. Therefore, we introduce the notion of \emph{exploitation paths} and \emph{attack nodes}. A conventional attack path chains a number of hosts from an entry to the final target, whereas an \emph{exploitation path} explicitly illustrates the product that is exploited at each host along with an attack path.
\emph{Attack nodes }then capture which product is chosen to exploit between a pair of connected hosts.

\begin{definition}[\textbf{Attack Nodes}]\label{def:att_nodes}
Given a network $N = \langle H, L, S, P \rangle$ with a specific product assignment $\as$,  $E = \{e_{\host{0}\host{1}}, \dots, e_{\host{m}\host{n}}\}$ denotes a set of attack nodes connecting a pair of connected hosts. Each attack node $e_{h_ih_j}$ includes a set of products on the destination host $h_j$, which can be exploited from a source host $h_i$, and a silent action (i.e. none). Therefore the domain states of an attack node is  $\Omega(e_{h_ih_j}) \in \as(\host{j}, \sh{j}) \cup \{none\} $.
\end{definition}
%The domain states of an attack node $e_{\host{i}\host{j}}$ include all assigned products to the destination host $\host{j}$, as well as a silent action \emph{none} denoting that no
%attack is performed at this step.
Attackers can choose one of the products to exploit or keep silent. Different choices lead to different propagating rates to the destination host. %In this way, a complete \emph{exploitation path} is a sequence of products to exploit from a root host which is the initial foothold acquired by attackers, to a target host of certain interests to attackers. %For a given network with assigned products, we can enumerate a set of possible exploitation paths.

\begin{figure}[!t]
  \centering
  \includegraphics[width=0.8\textwidth]{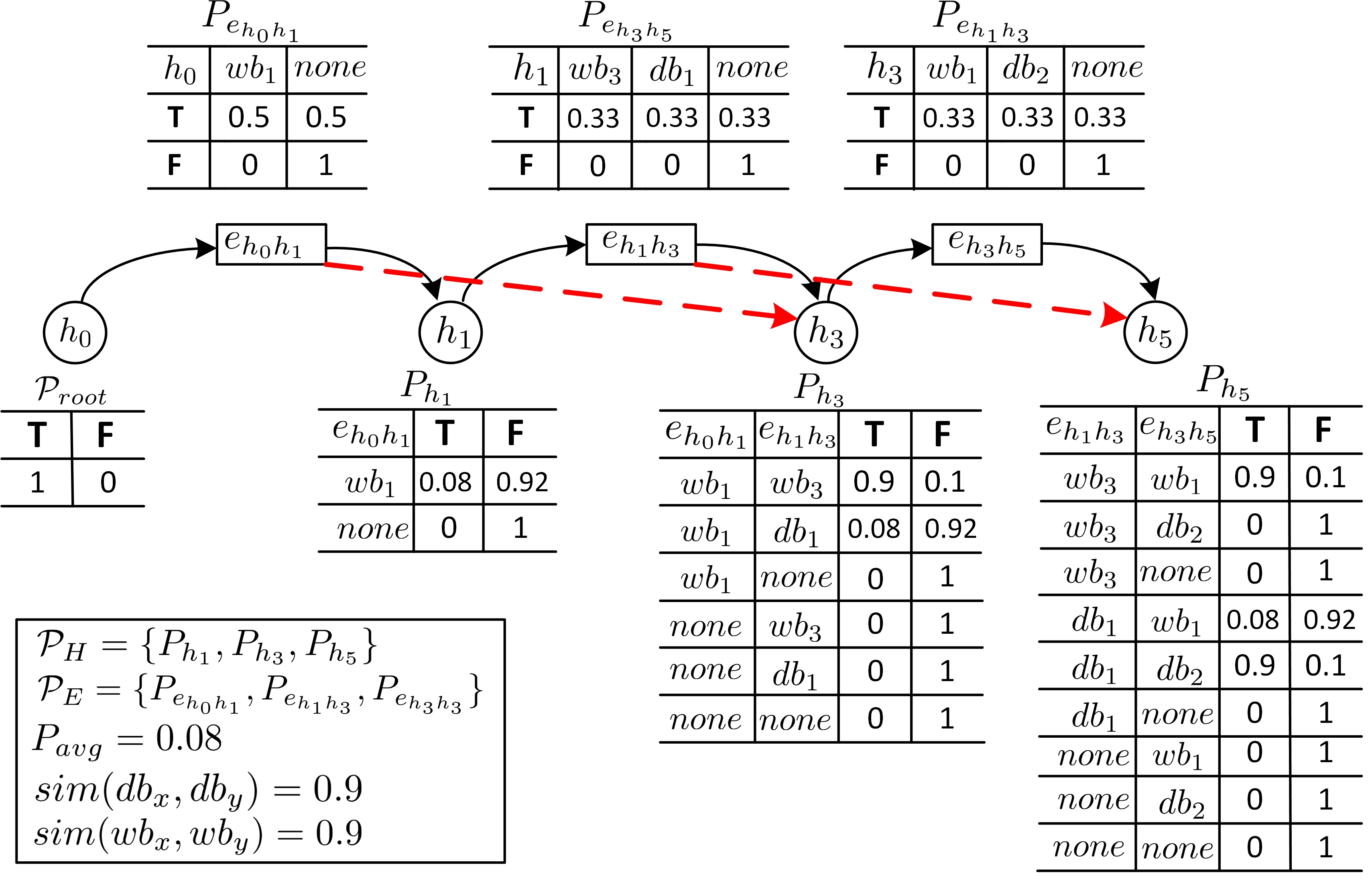}\\
  \caption{Partial Bayesian Network  for the running example }\label{fig:toy_bn}
\end{figure}

Now we can formally define the Bayesian Network (BN) for a given network $N$ subject to a specific assignment $\as$. Attack nodes $E$ are added into the BN. $\mathcal{P}_E $ defines for each attack node the likelihood of a product being selected to exploit. For instance, $\mathcal{P}_E = \{P_{e_{h_0h_1}}, P_{e_{h_1h_3}}, P_{e_{h_3h_3}}\}$ defines the conditional probability tables (CPT) for attack nodes in Figure \ref{fig:toy_bn} and in this example attackers choose products to exploit \emph{uniformly}. ~$\mathcal{P}'_{H} = \{P'_{h_1}, P'_{h_3}, P'_{h_5}\}$ defines the risk distribution of each host \emph{without} considering the vulnerability similarities of products, i.e. products share no vulnerabilities with each other, while $\mathcal{P}_{H} = \{P_{h_1}, P_{h_3}, P_{h_5}\}$ takes the similarities into account. Therefore, $\mathcal{P}'_{H}$ is constant for a given static network, regardless of the assigned products, while $\mathcal{P}_{H}$ is directly influenced by $\as$.

\begin{definition}\label{def:network}
Let $B = \langle \langle N, \as \rangle , E,  \mathcal{P}_{root}, P_{avg}, \mathcal{P}_E, \mathcal{P}_H, \mathcal{P}'_H \rangle$ be a \textbf{Bayesian Network}  for a given network $N = \langle H, L, S, P \rangle$ with a specific product assignment $\as$, where
\begin{itemize}
  \item $E$ is the associated set of attack nodes.
  \item $\mathcal{P}_{root}$ is the prior probability distribution of root hosts.
  \item $P_{avg}$ is the average infection rate of a zero-day exploit.
  \item  $\mathcal{P}_E = \{P_{e_{h_0h_1}}, \dots, P_{e_{h_mh_n}}\}$ includes conditional probability distribution (CPT) for all attack nodes such that $P_{e_{h_jh_i}}$ denotes  $Pr(e_{h_jh_i}~|~h_i)$, the probability distribution over a set of products to exploit next.
  \item $\mathcal{P}'_{H} = \{P'_{h_1}, \dots, P'_{h_n}\}$ includes conditional probabilities of all non-root host nodes given their preceding attack nodes. $P'_{h_k = T}$~denotes~$ Pr(h_k = T~|~\bigcup_{h_j\in H} e_{h_jh_k})$, and subject to noisy-OR operator \cite{pearl2014probabilistic},
       \begin{equation*}
          Pr(h_k = T~|~\bigcup_{h_j\in H} e_{h_jh_k})= 1- \prod_{ h_j\in H} 1-Pr(h_k = T | e_{h_jh_k})
       \end{equation*}
      where $P(h_k = T | e_{h_jh_k})$ is the probability of the host $h_k$ being compromised from $h_j$.
  \item $\mathcal{P}_{H} = \{P_{h_1}, \dots, P_{h_n}\}$ includes conditional probabilities of all non-root host nodes by considering the vulnerability similarity between products. $P_{h_k=T}$~denotes~$ Pr(h_k = T~|~\bigcup_{h_j,h_i\in H} e_{h_jh_k}, e_{h_ih_j})$, and subject to noisy-OR operator,
  \begin{eqnarray*}
     && Pr(h_k =  T ~|~\bigcup_{h_j, h_i\in H} e_{h_jh_k}, e_{h_ih_j})  ~~~~ ~~~~  ~~~~  \\
       && ~~~~~~~~~~~~~= 1- \prod_{ h_j, h_i\in H} 1-Pr(h_k = T |e_{h_jh_k}, e_{h_ih_j})
  \end{eqnarray*}
%where $Pr(h_k = T | e_{h_ih_j}, e_{h_jh_k})$ is the probability of the host $h_k$ being compromised through the preceding hosts $h_j$ and $h_i$, and
where $Pr(h_k = T | e_{h_ih_j}, e_{h_jh_k})$ is the probability of the host $h_k$ being compromised by considering the exploited products at the preceding hosts $h_j$ and $h_i$, and the probability is estimated as follows:
      \begin{eqnarray*}
        Pr(h_k = T ~|~ e_{h_ih_j} = \p{m}{j}, e_{h_jh_k} = \p{n}{k} ) \\
               = \begin{cases}
             \displaystyle P_{avg} & \text{if}~s_m \neq s_n \\
            \displaystyle sim(\p{m}{j}, \p{n}{k})  &   \scriptsize otherwise
          \end{cases}
      \end{eqnarray*}
\end{itemize}
\normalsize
\end{definition}

%\tlnote{needs reconsider the notation}

Without considering the similarity, the probability of a host being infected $P'_{h_k=T}$ only depends on the products being chosen to exploit at the host and the infection rate is set to the average zero-day propagation rate $P_{avg}$. To model the scenario of reusing exploits on similar products, we introduce an extra set of links into the BN, which are indicated by red dashed lines in Figure \ref{fig:toy_bn}. These links connect the preceding attack nodes with the current host such that we can represent  $P_{h_k}$. As demonstrated in Figure \ref{fig:toy_bn}, the probability distribution of $h_3$ is conditional on the current attack node $e_{h_1h_3}$ as well as its preceding attack node $e_{h_0h_1}$. If both attack nodes exploit the same type of products such as $wb_1$ and $wb_3$, then the chance of $h_3$ being compromised is the vulnerability similarity of $wb_1$ and $wb_3$, which is assumed to be 0.9. If different types of products are exploited, such as $wb_1$ and $db_1$, then the $P_{avg} = 0.08$ is used. Here we use the same value for $P_{avg}$ as in the existing work \cite{zhang2016network}\cite{wang2014modeling}. It is a nominal value but reasonably low for zero-day vulnerabilities, which is subject to change, depending on the assessment of the actual application scenarios.

From this simple example, we can see that different assignments would yield BN models with different infection rates on edges. Given a product assignment, we can construct the corresponding BN model to estimate the risk of the target node, which can be used to evaluate the current diversification introduced by the given assignment.

\subsection{Network Diversity Metrics}\label{sec:div}
In this subsection, we present the diversity metric used in this paper to evaluate  product assignments for a network. The network diversity metric was proposed by Zhang \emph{et. al.} \cite{zhang2016network} to evaluate a diversified network by measuring the \emph{average} attacking effort needed to compromise the network. We adapt the metric to fit our model considering the vulnerability similarity of products.

\begin{comment}
\begin{definition}[\textbf{Network Diversity Metrics $d_{min}$ and $d_{avg}$}]\label{def:d2}
Given a Bayesian Network $B = \langle \langle N, \as \rangle , E,  \mathcal{P}_{root}, P_{avg}, \mathcal{P}_E, \mathcal{P}_H, \mathcal{P}'_H \rangle$ constructed for a network $N$, a set of exploitation paths $\Gamma(\host{s},\host{t})$ can be generated  for an arbitrary root host $\host{s}$ and a target host $\host{t}$, hence the network diversity can be defined as below in term of the least attacking effort:
 \begin{equation}\label{eq:d2}
  d_{min} = \frac{ min_{l \in \Gamma(h_s, h_t)} D(l) }{  min_{l' \in \Gamma(h_s, h_t)} |l|-1   }
 \end{equation}

For a specific target host $h_t \in H$ the network diversity based on $B$ can be defined as below in term of the probability of the target host being compromised:
 \begin{equation}\label{eq:d2}
  d_{bn} = \frac{ P'_{h_t} }{ P_{h_t}}
 \end{equation}
\end{definition}
\end{comment}

\begin{definition}[\textbf{BN-based Diversity Metric  $d_{bn}$}]\label{def:d2}
Given a Bayesian Network $B = \langle \langle N, \as \rangle , E, \\ \mathcal{P}_{root}, P_{avg}, \mathcal{P}_E, \mathcal{P}_H,   \mathcal{P}'_H \rangle$ constructed for a diversified network $N$, and a specific target host $\host{t}$, the network diversity based on $B$ can be defined as below in term of the probability of the target host being compromised: $d_{bn} = \frac{ P'_{h_t=T} }{ P_{h_t=T}}$
 %\begin{equation}\label{eq:d2}
 % d_{bn} = \frac{ P'_{h_t=T} }{ P_{h_t=T}}
%\end{equation}
where $P_{h_t=T}$ ($P'_{h_t=T}$) is the probability of $h_t$ being infected with (without) considering the vulnerability similarity of products \emph{w.r.t }Definition \ref{def:network}.
\end{definition}

The probabilistic metric $d_{bn}$ estimates the average attacking effort by combining all valid exploitation paths. Naturally, the diversity metric $d_{bn}$ is always less than 1.0 and the greater value indicates higher diversity. With the help of Bayesian Networks, $P_{h_t=T}$ captures the risk of the target host when the vulnerability similarity of products is considered. $P_{h_t=T}$ reflects the \emph{current} robustness of the network, which is provided by the given product assignment, against repeating uses of zero-day exploits.  $P'_{h_t=T}$ indicates the \emph{maximum} potential of the network diversity. More explanations about this metric can be found in \cite{zhang2016network}.

%We use the running example in Figure \ref{fig:toy} to illustrate the calculation of the diversity metric $d_{bn}$. We assume $P_{avg} = 0.08$ and the similarity for any pair of products is 0.9. Given the assignment indicated by red circles,  $h_0$ is the attack entry point and $h_5$ is the target, we can obtain that  $d_{bn} =   \frac{ P'_{h_{5}=T} }{ P_{h_{5}=T}} =  0.01048/0.044630 = 0.235$ . \tlnote{re-calculate with avg = 0.08}

%\section{Formal Model of Network Diversity}

%\subsection{Product Assignment for a Network}\label{sec:pag}

%\section{Optimal Assignment of Diverse Products}\label{sec:opt}
\section{Case Study - Upgrading Legacy ICS with Modern Industrial Networks}
\label{sec:stuxnet}
%\tlnote{completely rewritten section}
In this section, we present a case study on upgrading legacy control systems with interconnected IT systems, to achieve the convergence of IT and OT in modern industrial networks. Such an integration can facilitate highly interconnected Industrial Internet of Things (IIoT) applications, but also leave ICS more vulnerable by introducing more attack vectors, i.e. as the control networks are no longer isolated, malware can propagate itself across IT systems to breach the core control units causing physical damage.

\begin{figure*}[!t]
  \centering
  \includegraphics[width=\textwidth]{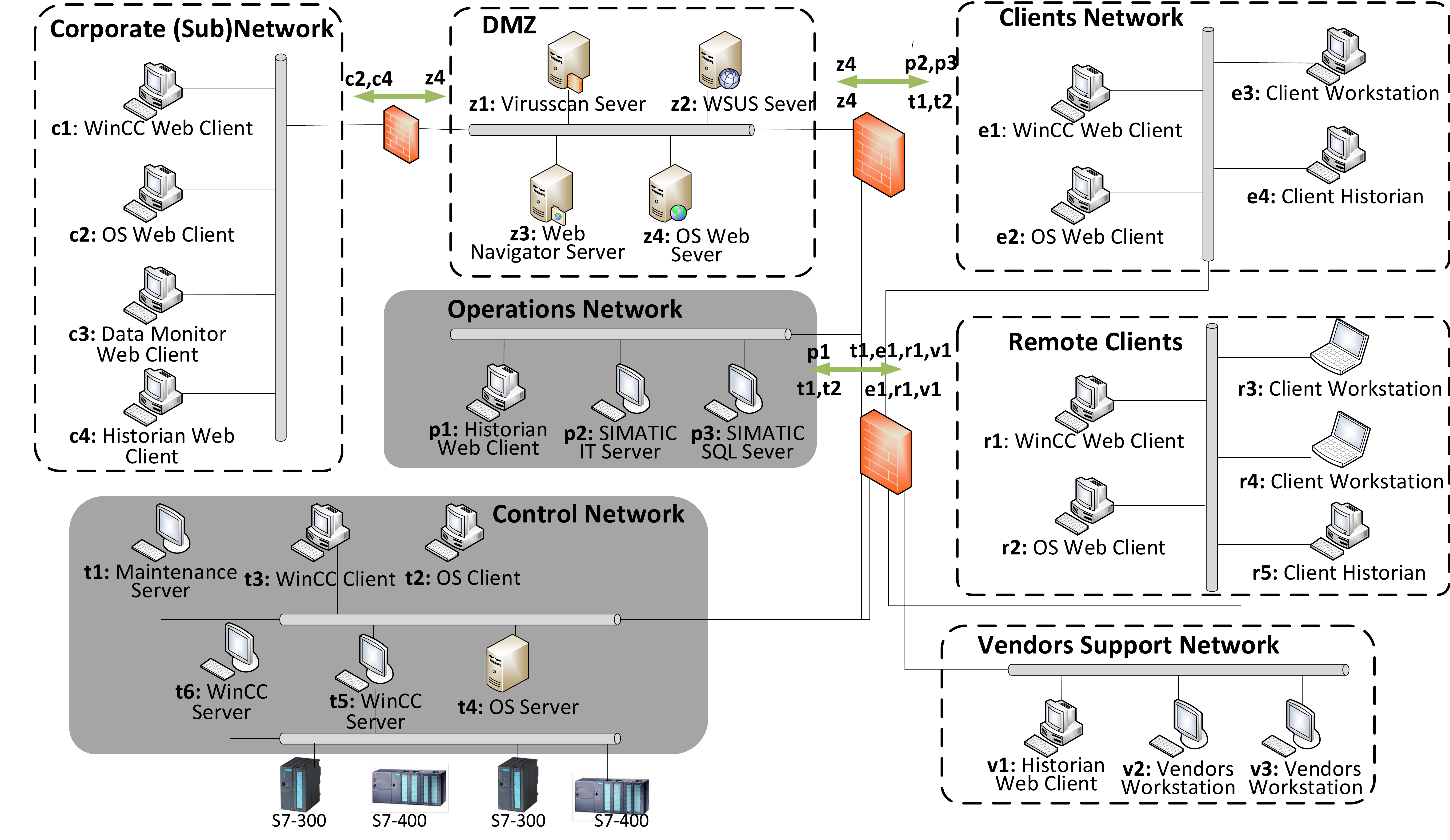}\\
  \caption{A typical structure of integrated modern ICS }\label{fig:case}
\end{figure*}

Therefore, in this case study, we demonstrate the usage of our approach in finding an optimal diversification strategy to improve the resilience of the integrated systems. Particularly we consider three main constraints that might arise when applying the approach in practice:
\begin{enumerate}[(i)]
\item Most hosts in OT networks run legacy software, which have \emph{no} flexibility to diversify or upgrade.
\item Some hosts in various networks are required to run specific software and hence cannot be diversified.
\item Some desirable and undesirable product combinations should be taken into account.
\end{enumerate}

We start with a brief description of the case study in Section \ref{sec:settings}. An optimal solution and two constrained optimal solutions are then computed and illustrated in Section \ref{sec:opt-solution}. In Section \ref{sec:case_eval} we evaluated the produced optimal solutions in terms of
\begin{inparaenum}[(i)]
\item the \emph{diversity metric} discussed in Section \ref{sec:metric}, and
\item the \emph{Mean-time-to-compromise (MTTC)} obtained from NetLogo simulations we have developed.
\end{inparaenum}

\subsection{Experiment Configuration}\label{sec:settings}
The example is adapted from the Stuxnet-like worm propagation analysis in \cite{stuxnet-prop}. Figure \ref{fig:case} depicts a typical ICS architecture integrating \emph{existing} OT zones (e.g. \emph{Operation Network, Control Network}) with \emph{new} IT zones (e.g. \emph{Corporate Sub-Network, Clients Network, Vendors Support Network}). We use gray shading to indicate that OT zones have \emph{no} flexibility to diversify or upgrade deployed software. Specific firewall white-list access rules are also given in Figure \ref{fig:case} to provide perimeter protection between different zones.

We use the example to demonstrate the Stuxnet worm propagation across an ICS. The primary intrusion can be introduced from the \emph{Corporate Network, Clients Network} or \emph{Vendors Support Network}. Once a host has been exploited as a foothold, the worm can continue scanning other connected hosts for similar vulnerabilities, by which the worm can propagate itself through the network. Stuxnet eventually breached the hosts in \emph{Control Network}, such as \textsf{t4},  \textsf{t5} and  \textsf{t6} in Figure \ref{fig:case}, which can access field devices.

In the following experiments, we consider the optimal assignment of products to provide three key services, i.e. an \emph{Operating System (OS)}, a \emph{Web Browser(WB)} and a \emph{Database Server(DB)}. These services are distributed across all the hosts in the network according to the key role each host plays (indicated in Figure \ref{fig:case}). For instance, the host \textsf{c1} in the \emph{Corporate Network} is configured as a WinCC Web Client, which runs WinCC V7.x as the main application. The essential requirements for this application are a Windows OS and an IE web browser \cite{wincc}, and hence a range of available products that we can choose to install on \textsf{c1} is provided in Table \ref{tab:products}. The host \textsf{z2} in DMZ requires a Windows OS and a Microsoft Database Server to run the WSUS server, which is reflected accordingly in the table. As a result, Table \ref{tab:products} lists essential services required at each host and the corresponding selections of products for each service.

\begin{table*}[!ht]
\centering
\caption{Available products for essential services in the case study }
\label{tab:products}
\scalebox{0.55}{
\begin{tabular}{|c|c|C{0.4cm}|C{0.4cm}|C{0.4cm}|C{0.4cm}|C{0.4cm}|C{0.4cm}|C{0.4cm}|C{0.4cm}|C{0.4cm}|C{0.4cm}|C{0.4cm}|C{0.4cm}|C{0.4cm}|C{0.4cm}|C{0.4cm}|C{0.4cm}|C{0.4cm}|C{0.4cm}|C{0.4cm}|C{0.4cm}|C{0.4cm}|C{0.4cm}|C{0.4cm}|C{0.4cm}|C{0.4cm}|C{0.4cm}|C{0.4cm}|C{0.4cm}|C{0.4cm}|}
\hline
\textbf{Serv.}  &  \textbf{Products}  & c1  & c2  & c3 & c4  & z1 & z2 & z3 & z4 &  \cellcolor{gray!40}p1 & \cellcolor{gray!40}p2 &  \cellcolor{gray!40}p3 &  \cellcolor{gray!40}t1 &  \cellcolor{gray!40}t2 &  \cellcolor{gray!40}t3 &  \cellcolor{gray!40}t4 &  \cellcolor{gray!40}t5 &  \cellcolor{gray!40}t6 &  e1 & e2 & e3 & e4 & r1&r2&r3&r4&r5& v1&v2&v3 \\ \hline \hline
\multirow{4}{*}{$\s{1}$}
% windows8
& Windows XP    & &  &  &  & & & & & \cellcolor{gray!40}& \cellcolor{gray!40}\checkmark& \cellcolor{gray!40}\checkmark& \cellcolor{gray!40}\checkmark& \cellcolor{gray!40}& \cellcolor{gray!40}\checkmark& \cellcolor{gray!40}\checkmark& \cellcolor{gray!40}\checkmark& \cellcolor{gray!40}\checkmark&  & & &  & & & & & &  & &\\ \cline{2-31}
% windows10
& Windows 7    & \checkmark & \checkmark & \checkmark & \checkmark & \checkmark &\checkmark& \checkmark& \cellcolor{gray!40} \checkmark&\cellcolor{gray!40} &\cellcolor{gray!40}&\cellcolor{gray!40}&\cellcolor{gray!40}&\cellcolor{gray!40}&\cellcolor{gray!40}&\cellcolor{gray!40}&\cellcolor{gray!40}&\cellcolor{gray!40} &\cellcolor{gray!40}\checkmark &\checkmark&\checkmark&\checkmark&\cellcolor{gray!40}\checkmark& \checkmark&\checkmark&\checkmark&\checkmark&\cellcolor{gray!40}\checkmark&\checkmark&\checkmark\\ \cline{2-31}
% ubuntu 14
& Ubuntu 14.04 & & \checkmark &  & \checkmark &\checkmark&    &    & \checkmark& \cellcolor{gray!40}\checkmark& \cellcolor{gray!40}   &  \cellcolor{gray!40}  & \cellcolor{gray!40}   &\cellcolor{gray!40}\checkmark& \cellcolor{gray!40}   &  \cellcolor{gray!40}  &  \cellcolor{gray!40}  &  \cellcolor{gray!40} &\checkmark &\checkmark &\checkmark &\checkmark& \checkmark&\checkmark &\checkmark &\checkmark &\checkmark & \checkmark &\checkmark &\checkmark \\ \cline{2-31}
% debian 8
& Debian 8.0   &            &  \checkmark    &       & \checkmark &  \checkmark  &    &    &  \checkmark &  \cellcolor{gray!40}  & \cellcolor{gray!40}   & \cellcolor{gray!40}   & \cellcolor{gray!40}   &\cellcolor{gray!40} &  \cellcolor{gray!40}  &  \cellcolor{gray!40}  & \cellcolor{gray!40}   &  \cellcolor{gray!40} & \checkmark & &\checkmark & \checkmark & \checkmark& &\checkmark &\checkmark &\checkmark & \checkmark&\checkmark & \checkmark\\ \hline \hline
\multirow{3}{*}{$\s{2}$}
& IE8          &  &  &  &  &    &    &&   & \cellcolor{gray!40}  &\cellcolor{gray!40}\checkmark& \cellcolor{gray!40}   &\cellcolor{gray!40}\checkmark&\cellcolor{gray!40}  &\cellcolor{gray!40}\checkmark&\cellcolor{gray!40}\checkmark&\cellcolor{gray!40}\checkmark&\cellcolor{gray!40}  \checkmark & \cellcolor{gray!40}  \checkmark & & &  &\cellcolor{gray!40}\checkmark &  & & & & \cellcolor{gray!40}\checkmark& &  \\ \cline{2-31}
& IE10         & \checkmark & \checkmark & \checkmark & \checkmark &    &    &\checkmark&\checkmark& \cellcolor{gray!40}&\cellcolor{gray!40}& \cellcolor{gray!40}   &\cellcolor{gray!40}& \cellcolor{gray!40} &\cellcolor{gray!40} &\cellcolor{gray!40}&\cellcolor{gray!40}&\cellcolor{gray!40} &   \checkmark & \checkmark &\checkmark & & \checkmark &\checkmark &\checkmark &\checkmark & & \checkmark &\checkmark &\checkmark\\ \cline{2-31}
& Chrome 50    &            & \checkmark &            &\checkmark&    &    &    & \cellcolor{gray!40}\checkmark&\cellcolor{gray!40}\checkmark& \cellcolor{gray!40}   & \cellcolor{gray!40}   & \cellcolor{gray!40}   &\cellcolor{gray!40}\checkmark&  \cellcolor{gray!40}  &  \cellcolor{gray!40}  &  \cellcolor{gray!40}  &  \cellcolor{gray!40} &\checkmark & &\checkmark&  & \checkmark&\checkmark &\checkmark &\checkmark & & \checkmark &\checkmark & \checkmark \\ \hline \hline
\multirow{4}{*}{$\s{3}$}
& MS SQL 2008  &            &            &   &   &    & &    &  & \cellcolor{gray!40} &\cellcolor{gray!40}\checkmark &\cellcolor{gray!40}\checkmark&\cellcolor{gray!40}\checkmark& \cellcolor{gray!40}   &\cellcolor{gray!40}\checkmark& \cellcolor{gray!40}   & \cellcolor{gray!40}   &\cellcolor{gray!40}   &  & & & & & & & & & & &\\ \cline{2-31}
& MS SQL 2014  &            &            & \checkmark &\checkmark  &    &\checkmark &    &   & \cellcolor{gray!40} &\cellcolor{gray!40}  & \cellcolor{gray!40}&\cellcolor{gray!40} &  \cellcolor{gray!40}  &\cellcolor{gray!40} & \cellcolor{gray!40}   &  \cellcolor{gray!40}  &  \cellcolor{gray!40}  & \cellcolor{gray!40} \checkmark & & & \checkmark &\cellcolor{gray!40}\checkmark & & & & \checkmark&\cellcolor{gray!40}\checkmark & &\\ \cline{2-31}
& MySQL 5.5    &            &            &    &\checkmark  &    &    &    &  & \cellcolor{gray!40} \checkmark&   \cellcolor{gray!40} &  \cellcolor{gray!40}  &  \cellcolor{gray!40}  &  \cellcolor{gray!40}  & \cellcolor{gray!40}   &  \cellcolor{gray!40}  &  \cellcolor{gray!40}  & \cellcolor{gray!40}   & \checkmark & & & \checkmark& \checkmark & & & & \checkmark&\checkmark& &\\ \cline{2-31}
& MariaDB 10   &            &            &            &\checkmark  &    &    &    &   &\cellcolor{gray!40} &  \cellcolor{gray!40}  &   \cellcolor{gray!40} &  \cellcolor{gray!40}  & \cellcolor{gray!40}   &  \cellcolor{gray!40}  &  \cellcolor{gray!40}  &  \cellcolor{gray!40}  & \cellcolor{gray!40}   &  \checkmark & & & \checkmark& \checkmark& & & & \checkmark & \checkmark& &\\ \hline
\end{tabular}}
\vspace{-3mm}
\end{table*} 

We highlight the legacy hosts in grey in Table \ref{tab:products}, which run outdated software and cannot be diversified (e.g. the host \textsf{p2}, \textsf{p3} in the \emph{Operations Network}). The example also includes several outdated versions of software running on legacy hosts such as Windows XP, MS SQL 2008. All of these introduce extra constraints in finding the optimal diversification strategy. The other chosen products in Table \ref{tab:products} are either frequently suggested in WinCC manuals or rated as one of the most vulnerable products by \emph{CVE Details} \cite{cve-details1}.

%In the following experiments, we consider a very low effectiveness of all firewalls in mitigating malware spread, because
%\begin{inparaenum}[(i)]
%\item we investigate the zero-day exploits and the defence offered by firewalls against such unknown attacks are very limited and negligible in our experiments, and
%\item our primary intention is finding a way to take maximum advantage of diversification to mitigate malware spread, and hence we keep the influence of other defensive factors (e.g. firewalls, access controls) to a minimum.
%\end{inparaenum}

The  similarities of web browsers and operating systems refer to Table \ref{tab:os-vuln} and \ref{tab:wb-vuln}, and the similarities for DB products are obtained in the same
way as described in Section \ref{sec:sim}. Given the products for each host in Table \ref{tab:products}, we can compute the optimal solution to diversifying the
networked ICS by the approach discussed in Section \ref{sec:opt}. It is worth noticing that our approach offers high generality and flexibility, by which each host
can have a customized range of services, and each service can have various ranges of products to deploy.% at different hosts subject to any compatibility requirementson the hosts.

\subsection{Optimal Assignment of Products}\label{sec:opt-solution}

\begin{figure*}[!t]
  \centering
  \includegraphics[width=0.5\textwidth]{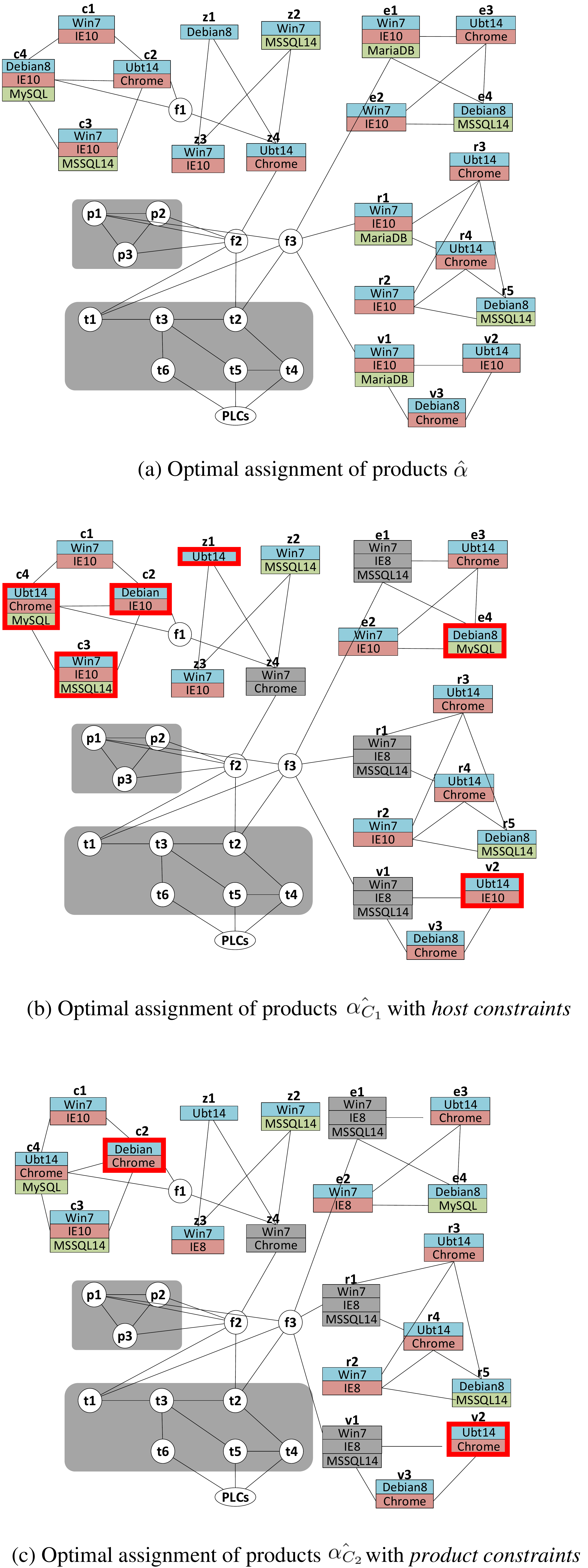}\\
  \caption{Optimal Assignment of products for the case study }\label{fig:opt}
\end{figure*}

The optimal assignment $\opm{\as}$ for the case study is computed by the approach introduced in Section \ref{sec:opt} and illustrated in Figure \ref{fig:opt}(a). The assignment indicates the optimal strategy to deploy the software in IT networks when integrating with OT systems. % in Operations Network and Control Network.
%Different colors of boxes correspond to different services of each host.
The solution attempts to minimize the vulnerability similarity between each pair of connected hosts. From the figure, we can find that each pair of connected hosts is generally assigned with different products from each other.

%Such locally optimal assignments might be adjusted by our optimizer to pursue the global optimum. For instance, the optimal solution suggests  \emph{Chrome 50} for both ends of the edge (\textsf{c4},\textsf{c2}). This is because \textsf{c1} and \textsf{c3} have limited choices of web browsers (i.e. only IE 8 and IE 10 available), and diversifying the web browser on \textsf{c2} (or \textsf{c4}) to IE would actually lead to  highly similar products on two more edges (\textsf{c1}, \textsf{c2}) and (\textsf{c3}, \textsf{c2}) (or  (\textsf{c1}, \textsf{c4}) and (\textsf{c3}, \textsf{c4})). Therefore, our system made the most optimal decisions on this. %\tlnote{not sure if I explained this clearly}

%\tlnote{change IE8 to IE10 in fig6.}
%\tlnote{IE8 is only for legacy hosts in Table 5}

As mentioned in the beginning of this section, the second type of constraints we might encounter is that some hosts are required to run specific software according to certain company policies. For this case study, we specify that the host $\textsf{z4}$, \textsf{e1}, \textsf{r1} and \textsf{v1} are required to run specific products. We outline fixed choices for these hosts in Table \ref{tab:products} in grey. Having adding those constraints into the optimisation, we now compute the constrained optimal assignment $\opm{\as}_{C_1}$, which is given in Figure \ref{fig:opt}(b). It can be seen that whilst we fixed the products of the four hosts, the new solution accordingly updates assignments of products for several hosts to find new optimal diversification, as highlighted by red squares.

We can also specify undesirable product combinations to avoid during optimisation. For instance, the solution $\opm{\as}_{C_1}$ in Figure \ref{fig:opt}(b) uses the IE10 on Ubuntu14.04 at host \textsf{v2}. If we want to eliminate such undesirable assignments, we can specify and embed \emph{product constraints} in the computation of optimal solutions, as introduced in Definition \ref{def:constraints}.

For instance, the following set of \emph{global} constraints $\mathcal{C}_2 = \{c_1, c_2, c_3, c_4\}$ captures the exclusive requirement between Ubuntu (Debian) OS and IE across all hosts:

\begin{align*}
  c_1 :=& <\textsf{ALL},~ \emph{\text{OS},~ \text{WB},~ +\text{Ubuntu 14.04},~ -\text{IE8}} > \\
    c_2 :=& <\textsf{ALL},~ \emph{\text{OS},~ \text{WB},~ +\text{Ubuntu 14.04},~ -\text{IE10}} > \\
  c_3 :=& <\textsf{ALL},~ \emph{\text{OS},~ \text{WB},~ +\text{Debian 8.0},~ -\text{IE8}} > \\
    c_4 :=& <\textsf{ALL},~ \emph{\text{OS},~ \text{WB},~ +\text{Debian 8.0},~ -\text{IE10}} >
\end{align*}

With the constraints in $\mathcal{C}_2$, we can compute the constrained optimal solution $\opm{\as}_{C_2}$ that is illustrated in Figure \ref{fig:opt}(c). It can be found that the web browsers at \textsf{c2} and \textsf{v2} are changed to \emph{Chrome} as required. %However, the solution $\opm{\as}_{C_1}$ also introduces undesirable product combinations at \textsf{c2} and \textsf{t2} by assigning IE8 and IE10 respectively.

%Therefore, in another run of the optimization, we define a set of global constraints $\mathcal{C}_2 = \{c_3, c_4\}$ as below to obtain a new constrained optimal solution $\opm{\as}_{C_2}$.
%\begin{eqnarray*}
%  c_3 &:=& <\textsf{ALL},~ \emph{\text{OS},~ \text{WB},~ +\text{Ubuntu 14.04},~ -\text{IE8}} > \\
%    c_4 &:=& <\textsf{ALL},~ \emph{\text{OS},~ \text{WB},~ +\text{Ubuntu 14.04},~ -\text{IE10}} >
%\end{eqnarray*}
%We can visualize the assignment by $\opm{\as}_{C_2}$ in Figure \ref{fig:opt}(c), where all hosts now meet the specified constraints.

%\tlnote{ignore e1 and e2}
%\tlnote{explain both chromes on c2 and c4}

The optimal solution $\opm{\as}$ is produced by minimizing the energy function presented in Section \ref{sec:energy-opt}, and hence it guarantees the minimal infection rate of the worm and the most diverse product assignment possible. In order to accommodate the host and product constraints,  the constrained solutions  $\opm{\as}_{C_1}$ and  $\opm{\as}_{C_2}$ have to sacrifice a certain amount of diversity. In the next section, we evaluate all these optimal solutions and quantify the compromised diversity of the constrained solutions in terms of the diversity metric proposed in Section \ref{sec:div} and \emph{MTTC} by our NetLogo simulation.
%introduce an attack simulation  by using  NetLogo to mimic the penetration of  zero-day exploits.

\subsection{Case study analysis}\label{sec:case_eval}

\subsubsection{Evaluation by Network Diversity Metric}

First of all, we construct a Bayesian Network for the case study with a given assignment of products in order to estimate the propagation of malware. In the following experiments, we consider an attacker breaks into the system from \textsf{c4} in Corporate Network, and hence we set \textsf{c4} to be the root being infected with a prior probability 1.0.  The final target of the attack is set to the host \textsf{t5} which has the direct access to controlling the critical field devices. Therefore, the probability of the target \textsf{t5} being infected becomes the key element to calculate the network diversity metric $d_{bn}= P'_{t5=T}/ P_{t5=T}$, as defined in Definition \ref{def:d2}.

Given an assignment of products (e.g. the optimal one $\opm{\as}$), we can determine the possible infection rate of zero-day malware at each edge with the help of the constructed Bayesian Network. As we investigate the infection of multiple zero-day exploits, we assume that the attacker is in possession of three unique zero-day exploits, each of which exploits a particular type of product respectively (i.e. OS, WB and DB). Once a host is infected, attackers search for similar products/vulnerabilities to exploit amongst the connected hosts and proceed. When multiple exploits are feasible, attackers evenly choose one to use, which defines the CPT for attack nodes associated with each edge. The similarity between the source and chosen product decides the likelihood of infecting the chosen product. %Therefore, it can be seen that different assignments of products render different Bayesian Networks of malware propagation.

%\tlnote{todo: add description of fig6}
\begin{table}[!h]
\centering
\caption{Diversity metric $d_{bn}$ of different product assignments}
\label{tab:case_eval}
\scalebox{0.95}{
\begin{tabular}{|c|c|c|c|c|}
\hline
\textbf{Label} &  \textbf{Description}                                     & $\log P'_ {t5=T}$ & $\log P_{t5=T}$ &  \specialcell{$d_{bn}= \frac{ P'_{t5=T} }{ P_{t5=T}}$} \\ \hline\hline
$\hat{\alpha}$    & optimal assign.     & -3.151       & -3.062     & 0.81457            \\ \hline
 $\hat{\alpha}_{C_1}$   & host constr.  & -3.151       & -2.838      & 0.48590         \\ \hline
  $\hat{\alpha}_{C_2}$   &product constr.  & -3.151      & -2.833      & 0.48119         \\ \hline
   $\alpha_r$   & random assign.         & -3.151       & -2.576      & 0.26622           \\ \hline
    $\alpha_m$   & mono assign.              & -3.151       & -1.978      & 0.06709           \\ \hline
\end{tabular}}
\end{table}

%\tlnote{needs to calculate the results for   $\hat{\alpha}_{C_2}$ }

The first row of Table \ref{tab:case_eval} is the evaluation of the optimal assignment $\opm{\as}$ which reaches a very high diversity $d_{bn} = 0.81457$. The constrained optimal solutions  $\hat{\alpha}_{C_1}$ and  $\hat{\alpha}_{C_2}$  produce lower diversities as the two solutions are required to accommodate certain constraints. Discussion about the impact of such  constraints on the optimal diversification continues in Section \ref{sec:constraints}.

For the purpose of comparison, we also generate a homogeneous assignment $\as_m$ which generally allocates the \emph{same} operating system,  the \emph{same} web browser and the \emph{same} database server for all non-constrained hosts. Such mono-assignment provides the \emph{worst} possible diversity for the ICS case. It also shows how vulnerable the network would become if we use homogeneous products. Besides, a randomly diversified assignment $\as_r$ is also provided, which delivers a limited diversity that is significantly lower than our optimal solution.

The notation $P'_ {t5=T}$ denotes the probability of the target \textsf{t5} being infected \emph{without} considering the vulnerability similarities between products. Therefore,  $P'_ {t5=T}$  has a constant value for all different assignments. When we take similarities into consideration, the probability of  \textsf{t5} being infected $P'_ {t5=T}$ increases with less diverse assignments of products.

\subsubsection{Evaluation by NetLogo Simulation}
%\tlnote{completely reworked section}
%\tlnote{somewhere to mention As most cyber threats against ICS initiated from interconnected IT systems, we believe that more diverse and resilient deployment of products in the interconnected IT systems can significatively reduce the chance of malware propagation and protect the ICS. }

NetLogo is an agent-based modelling tool that enables a programmable modelling environment for simulating natural phenomena and behaviours of complex systems over time \cite{netlogo}. We use NetLogo to construct the networked ICS as shown in Figure \ref{fig:case} and simulate the propagation of malware. After breaking into the system from a host, attackers can further spread the worm across the network. Given an assignment of products (e.g. the one in Figure \ref{fig:opt}(a)), we can determine the possible infection rate of zero-day exploits at each edge in NetLogo.

\begin{figure*}[!t]
  \centering
  \includegraphics[width=\textwidth]{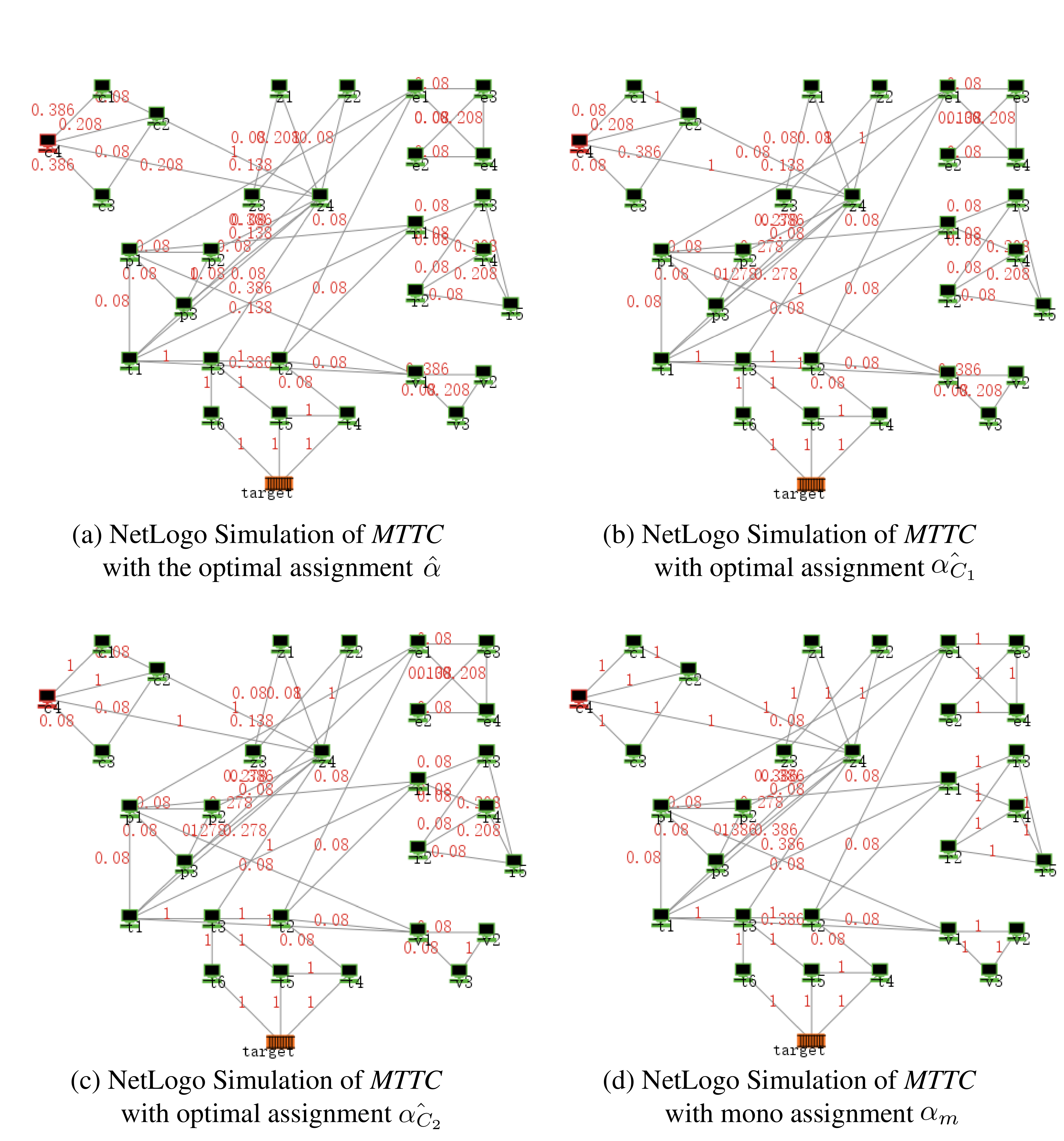}\\
  \caption{NetLogo simulation views for the case study }\label{fig:nl}
\end{figure*}

Figure \ref{fig:nl} provides the simulation views for four different assignments $\hat{\alpha}$,  $\hat{\alpha}_{C_1}$ ,  $\hat{\alpha}_{C_2}$ and $\alpha_m$ respectively. The numbers on edges are the highest possible infection rate of exploits between a pair of hosts. As we considered the sophisticated attackers who conduct reconnaissance activities before launching attacks, at each step attackers always chooses the exploits with the highest success rate. It can been seen from Figure \ref{fig:nl}(a) that the optimal assignment without constraints $\hat{\alpha}$ guarantees relatively low infection rate for most edges except those with legacy hosts. Due to rigid constraints, the $\hat{\alpha}_{C_1}$ and  $\hat{\alpha}_{C_2}$ have to leave the infection rates on some edges to be 1, as the connected hosts are assigned with the same product, e.g. \emph{IE10 }on \textsf{c1} and \textsf{c2}, and\emph{ Window 7} on \textsf{z2} and \textsf{z4} in Figure \ref{fig:nl}(b).

Having set up the simulation with a given product assignment, we can determine how much time (\emph{MTTC}) is required by attackers to penetrate the diversified network, which implies the average effort required to compromise the network. More optimal assignment should provide more resilience to the network against the penetration.

To test the resilience provided by the diversification, we designed five sets of experiment to simulate the malware propagation from five different entry points respectively -- \textsf{c1} and \textsf{c4} from the \emph{Corporate Network}, \textsf{e3} from the \emph{Clients Network}, \textsf{r4} from the \emph{Remote Clients}, and \textsf{v1} from the \emph{Vendors Support Network}. Once the entry host is infected, attackers search for similar products/vulnerablities to exploit from the connected hosts.

%\tlnote{new figure (Figure \ref{fig:nl}) is added}

\begin{table}[!h]
\centering
\caption{MTTC (in ticks) against different assignments}
\label{tab:case_mttc}
\scalebox{0.95}{
\begin{tabular}{|c|c|c|c|c|c|}
\hline
\textbf{Assignment} &  \specialcell{MTTC \\ from \textsf{\textbf{c1}}}  & \specialcell{MTTC \\ from \textsf{\textbf{c4}}} & \specialcell{MTTC \\ from \textsf{\textbf{e3}}} & \specialcell{MTTC \\ from \textsf{\textbf{r4}}} & \specialcell{MTTC \\ from \textsf{\textbf{v1}}} \\ \hline\hline
$\hat{\alpha}$    & 45.313    &  37.561       & 52.663     & 52.491    &   24.053      \\ \hline
 $\hat{\alpha}_{C_1}$   & 28.041  &16.812     & 44.359     & 48.472  &  15.243    \\ \hline
  $\hat{\alpha}_{C_2}$   & 14.549  & 15.817    & 45.118     & 46.257  & 14.749      \\ \hline
  $\alpha_{m}$   & 14.345  & 12.654    & 19.338    & 18.865  & 15.916     \\ \hline
\end{tabular}}
\end{table}

%\tlnote{consider to add the results for mono and random as well in table 7}
%\tlnote{add descrption of mono MTTC}
For each set of experiments, we deployed the network according to the three optimal assignments  $\hat{\alpha}$,  $\hat{\alpha}_{C_1}$  and  $\hat{\alpha}_{C_2}$, as well as the mono-assignment $\alpha_m$. Each experiment ran the simulation for 1,000 times. The average MTTC for each test is given in Table \ref{tab:case_mttc}.  The MTTC is the time steps (i.e. ticks in NetLogo) consumed by attackers to successfully reaching the final target. The results show that the optimal assignment $\hat{\alpha}$ provides the strongest resilience to the network, as it requires the longest period of time to be compromised in the all five attack scenarios, while the other two constrained optimal assignments can be compromised in a shorter period of time. The mono-assignment provides the weakest resilience to the network.

\section{Network Diversity Analysis}\label{sec:eval}
%\tlnote{needs a more accurate title for this section...}
%\tlnote{no changes made in this section}
In addition to the vulnerability similarity of products, there are other factors affecting the optimization of network diversity. In this section, we focus on three key factors  -- the network structure (Section \ref{sec:routing}), the variety of products per service (Section \ref{sec:products}) and the number of configuration constraints (Section \ref{sec:constraints}). We use an artificial network  $N$ in which 30 hosts $H = \{h_1, \dots, h_{30}\}$ are created, each host runs three different services $S = \{s_1, s_2, s_3\}$ and at least three products are provided for each service to choose. We set three entry nodes and a target $h_{30}$. The diversity metric is calculated in terms of the probability of $h_{30}$ being compromised. In the following sections, we create a number of variants of the network $N$ to examine the impact of the aforementioned factors on the optimal network diversity.

\subsection{Impact of Network Structure} \label{sec:routing}
We create different network structures with various numbers of routing nodes.  Routing nodes are generally hosts with heavy traffic flows. Diversifying routing nodes is of great importance to improve the network robustness \cite{newell2015increasing}. We define \emph{routing nodes} as the hosts of at least degree 3, i.e. there are at least three edges connecting to the host.  We create different numbers of routing nodes by randomly adding a number of edges to the network. The number of routing nodes in the following experiments are 6, 10, 12 and 14 respectively.

\begin{table}[!h]
\centering
\caption{Diversity metric $d_{bn}$ of different \# routing nodes}
\label{tab:routing}
\scalebox{0.95}{
\begin{tabular}{|c|c|c|c|c|}
\hline
 \specialcell{\textbf{\# Routing} \\ \textbf{ Nodes}} &  \specialcell{\textbf{\# Attack} \\ \textbf{Paths}}& $\log P'_{h_{30}=T}$ & $\log P_{h_{30}=T}$ & $d_{bn}=  \frac{ P'_{h_{30}=T} }{ P_{h_{30}=T}} $ \\ \hline\hline
6                         & 15                             & -3.896          & -3.814        & 0.82682                    \\ \hline
10                        & 48                             & -6.239          & -5.496         & 0.18095                     \\ \hline
12                        & 60                             & -6.238          & -5.495         & 0.18093                     \\ \hline
14                        & 66                             & -5.829          & -5.238         & 0.25628                     \\ \hline
\end{tabular}}
\end{table}

Table \ref{tab:routing} gives the evaluation of the optimal diversification for each case. At a low number of routing nodes (i.e. 6), the optimal diversification provides a remarkable improvement to network diversity with $d_{bn} = 0.82682$. As the increasing number of attack paths, the optimal diversity metric has been reduced. Despite the growing number of attack paths,  our optimal solutions can still maintain the similar network diversity at the 10, 12 and 14 routing nodes. It should be noticed that the increasing number of routing nodes are created by adding \emph{random} edges to the network. Therefore, the trend of the diversity $d_{bn}$ could be fluctuate as the number of routing nodes increases. The optimal solutions also improve the network robustness against the expanding attack vector, which is reflected by the generally decreasing probability of the target being infected $P_{h_{30}=T}$. Another noticeable observation is that the optimal diversification tends to be more necessary when protecting dense network with larger number of exploitation paths. %We also need to emphasize that the edges added to each case are randomly selected and hence the difference of $d_{bn}$ between one and another is not  xxx.

%\tlnote{double check \#  exploitation paths}

\subsection{Impact of the Variety of Products}\label{sec:products}

The variety of products can also influence the optimal  diversity. A wide variety of candidate products can introduce more diversity and flexibility when assigning products, which more importantly, can reduce the chance of a pair of connected nodes being assigned highly similar products. We still use the network $N$ of 30 hosts with 10 routing nodes, 3 services per host. The variety of products we tested is from 3 to 7 and the detailed evaluation is provided in Table \ref{tab:eval_products}.

\begin{table}[!h]
\centering
\caption{Diversity metric $d_{bn}$ of various \# products per service}
\label{tab:eval_products}
\scalebox{0.92}{
\begin{tabular}{|c|c|c|c|}
\hline
 \specialcell{\textbf{\# Products} \\ \textbf{ per Service}}  & $\log P'_{h_{30}=T}$ & $\log P_{h_{30}=T}$ & $d_{bn}=  \frac{ P'_{h_{30}=T} }{ P_{h_{30}=T}} $ \\ \hline\hline
3                                & -6.239          & -5.062         & 0.06653                     \\ \hline
4                                & -6.239          & -5.216         & 0.09481                     \\ \hline
5                                & -6.239          & -5.392         & 0.14226                     \\ \hline
6                                & -6.239          & -5.655         & 0.26053                     \\ \hline
7                                & -6.239          & -5.950         & 0.51437                     \\ \hline
\end{tabular}}
\end{table}

The decreasing value of  $P_{h_{30}=T}$ indicates that with a higher variety of products, the network can provide better protection to the target. We  also observe that the diversity $d_{bn}$ is  improved with more available products. However, when applying the optimal diversification in practice, as demonstrated by the ICS case study in Section \ref{sec:opt-solution}, a number of configuration constraints can stop us from using the most optimal assignment. In the next section, we study the impact of the increasing  number of  constraints on the optimization.

\subsection{Impact of Configuration Constraints}\label{sec:constraints}

We specify different numbers of \emph{local} constraints (i.e. 5, 10, 15) to compute the constrained optimal solution. The added constraints for each run of the experiment are randomly generated. It is possible that the added constraints are not in conflict with the optimal solution, in which case the resulting constrained optimal solution can still provide the same diversity as the optimal solution. Therefore, we intentionally add constraints to against the optimal solution generated in Table \ref{tab:routing} (at 10 routing nodes), so that we can evaluate how the constraints compromise the optimal diversity.

\begin{table}[!h]
\centering
\caption{Diversity metric $d_{bn}$ of various \# constraints}
\label{tab:constraints}
\scalebox{0.92}{
\begin{tabular}{|c|c|c|c|}
\hline
\textbf{\# Local Constraints} & $\log P'_{h_{30}=T}$ & $\log P_{h_{30}=T}$ & $d_{bn}=  \frac{ P'_{h_{30}=T} }{ P_{h_{30}=T}} $ \\ \hline\hline
0                                & -6.239          & -5.497         & 0.18095                     \\ \hline
5                                & -6.239          &  -5.490        & 0.17838                     \\ \hline
10                                & -6.239          & -5.443         &  0.15996                   \\ \hline
15                                & -6.239          &  -5.224        &  0.09669                   \\ \hline
\end{tabular}}
\end{table}

The results are then compared with the optimal solution with no constraint in Table \ref{tab:constraints}. The results clearly show that the increasing number of constraints can compromise the optimized diversity and reduce the protection to the target.

\section{Scalability Analysis}\label{sec:random}
\begin{figure*}[!t]
  \centering
  \includegraphics[width=\textwidth]{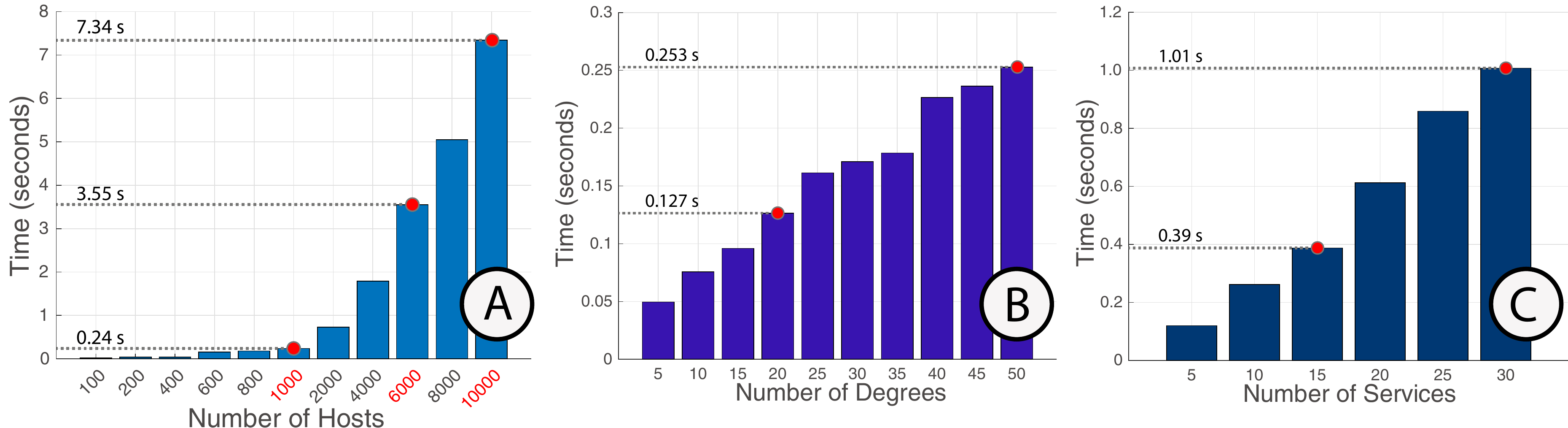}\\
  \caption{ Scalability with randomly generated networks. (A) analysis by fixing 3 degrees  and 3 services per host; (B) analysis by fixing 100 hosts and 3 services per host; (C) analysis by fixing 100 hosts and 30 degrees per host.}\label{fig:time}
\end{figure*}

%\tlnote{no changes made in this section}
In this section, we focus on analyzing the scalability of our optimization approach. We run the optimization against a series of randomly generated networks. Figure~\ref{fig:time} illustrates a numerical analysis when optimizing networks of varying numbers of hosts in Figure~\ref{fig:time}(A), varying degrees of hosts (edges per host) in Figure~\ref{fig:time}(B) and varying numbers of services per host in Figure~\ref{fig:time}(C).

\begin{table*}[!t]
\centering
\caption{Computational time (in seconds) for networks of various densities over different \# hosts}
\label{tab:hosts}
\scalebox{0.8}{
\begin{tabular}{|c|c|c||C{1cm}|C{1cm}|C{1cm}|C{0.9cm}|C{1cm}|C{1cm}|C{1cm}|C{1cm}|C{1.1cm}|}
\hline
\multirow{2}{*}{} & \multirow{2}{*}{\# deg.} & \multirow{2}{*}{\# serv.} & \multicolumn{9}{c|}{\# hosts }                                             \\ \cline{4-12}
                  &                             &                           & \textbf{100}  & \textbf{200}   & \textbf{400}   & \textbf{600}   & \textbf{800}   & \textbf{1000}   & \textbf{2000}   & \textbf{4000}   & \textbf{6000}    \\ \hline\hline
mid-density    & \textbf{20}                          & \textbf{15}                        & 0.239 & 0.438  & 1.099 & 1.478 & 1.944 & 2.784  & 6.706  & 16.517 & 33.392  \\ \hline
high-density      & \textbf{40}                          & \textbf{25}                        & 0.640 & 1.766 & 3.553 & 5.881 & 8.135 & 10.999 & 27.484 & 82.500 & 151.110 \\ \hline
\end{tabular}}
\end{table*}

\begin{table*}[!t]
\centering
\caption{Computational time (in seconds) for various sizes of networks over different \# degrees}
\label{tab:degrees}
\scalebox{0.7}{
\begin{tabular}{|c|c|c||c|c|c|c|c|c|c|c|c|c|}
\hline
\multirow{2}{*}{} & \multirow{2}{*}{\# hosts} & \multirow{2}{*}{\# serv.} & \multicolumn{10}{c|}{\# degree}                                                                                                          \\ \cline{4-13}
                  &                           &                              & \textbf{5} & \textbf{10} & \textbf{15} & \textbf{20} & \textbf{25} & \textbf{30} & \textbf{35} & \textbf{40} & \textbf{45} & \textbf{50} \\ \hline \hline
mid-scale         & \textbf{1000}             & \textbf{15}                  & 0.759       & 1.577       & 1.954       & 2.693       & 3.294       & 4.040       & 4.652        & 5.174       & 5.758       & 6.309       \\ \hline
large-scale       & \textbf{6000}             & \textbf{25}                  & 21.239     & 40.940      & 59.216      & 77.583      & 95.750      & 117.810     & 144.470     & 152.040     & 167.190     & 189.710     \\ \hline
\end{tabular}}
\end{table*}

\begin{table*}[!t]
\centering
\caption{Computational time (in seconds) for various sizes of networks over different \# services}
\label{tab:services}
\scalebox{0.84}{
\begin{tabular}{|c|c|c|c||c|C{1.2cm}|C{1.2cm}|C{1.2cm}|C{1.2cm}|C{1.2cm}|C{1.2cm}|}
\hline
\multirow{2}{*}{} & \multirow{2}{*}{\# hosts} & \multirow{2}{*}{\# deg.} & \multirow{2}{*}{\# edges} & \multicolumn{6}{c|}{\# services}                                                     \\ \cline{5-10}
                  &                           &                     &     & \textbf{5} & \textbf{10} & \textbf{15} & \textbf{20} & \textbf{25} & \textbf{30} \\ \hline \hline
mid-scale         & \textbf{1000}             & \textbf{20}         &  $\sim$ \textbf{20,000}   & 0.603       & 1.608       & 2.709       & 4.008       & 5.253       & 6.974       \\ \hline
large-scale       & \textbf{6000}             & \textbf{40}         &  $\sim$ \textbf{240,000}     & 10.306     & 27.214      & 51.587      & 90.407      & 134.340     & 188.050     \\ \hline
\end{tabular}}
\end{table*}

For the best performance, our optimizer is implemented using \texttt{C++} and enables the multi-threading mechanism to provide high convergence speed in multi-level optimization. We apply a GPU-friendly compute unified device architecture to gain extra efficiency on complex matrix operation. All the experiments run on a mid-range computer with an Intel i5 2.8GHz CPU, a 8GB RAM and an Nvidia GTX 750. The optimization in all the following experiments can be achieved within a reasonably short time from a couple of seconds to minutes.

We observe that the number of hosts has a major impact on the computational consumption. As shown in Figure~\ref{fig:time}(A), the time increases nonlinearly with the increasing number of hosts. However, our method still provides a high efficiency on mid-scale networks -- the optimization converges within 0.24 seconds when the network size is up to 1000 hosts. A reasonably high speed is also provided on the large-scale networks -- the optimal assignment for 10,000 hosts can be obtained within 7.342 seconds on average.

Figure~\ref{fig:time} (B) and (C) show that the computational time of our approach increases linearly along with the increment of either the degree of nodes or the number of services. By fixing the network size (\#100) and the number of services (\#3) of each host, our optimization converges within 0.253 seconds for the large degree (\#50) that results in more than 2900 edges. Our method behaves similarly on the experiments with varying numbers of services. The optimal solution is found within 1.01 seconds given a large number of services (\#30) on a network with 100 hosts. Such computational time is highly promising for most networks in the real world.

We further test our optimization approach against high density and large-scale networks that we might encounter in practice. Table \ref{tab:hosts} provides the computational time of optimizing networks with the middle (20 degrees and 15 services per host) and high density (40 degrees and 25 services per host). Again we observe that the number of hosts has a major impact on the computational time, but our method still finds the optimal solution within 3 minutes for large-scale (6000 hosts) high-density network.  Moreover, we run experiments on mid-scale and large-scale networks with various densities and the results in Table \ref{tab:degrees} show that the degree has less influence on the computational time than the number of hosts. Finally, we vary the number of services for each host on both mid-scale and large-scale networks  in Table \ref{tab:services}. For a large-scale network of 6000 hosts with up to 240,000 edges and 30 services per host, our method still performs well and converges at about 3 minutes.

\section{Discussion and Conclusion}\label{sec:conclusion}
%\tlnote{updated section}

%\tlnote{somewhere to mention As most cyber threats against ICS initiated from interconnected IT systems, we believe that more diverse and resilient deployment of products in the interconnected IT systems can significatively reduce the chance of malware propagation and protect the ICS. }

Moving towards integrated ICS enables an efficient way to operate these systems, but also provides new attack vectors for adversaries to breach them. It is now a challenging and urgent issue for many industrial organizations to find a secure way to converge OT and IT systems to provide an efficient and also resilient industrial environment. Furthermore, there are other constraints hindering us from finding an optimal solution, such as outdated legacy systems, strict company policies and other configuration requirements. In this paper, we proposed an approach based on software diversification to increase the system resilience of the integrated ICS against malware propagation.

We introduced the similarity metric to capture how similar the vulnerabilities of two products are, which was then applied in a statistical study on CVE/NVD databases. The study showed that most vulnerabilities could affect multiple products, even from different vendors. Therefore, when finding the diverse assignment of products, we explicitly considered such vulnerability similarities of products. The similarity metric can estimate the likelihood of a zero-day exploit successfully propagating itself between two products.  By assigning diverse products for a pair of connected hosts, such propagation can be effectively reduced. Unlike most existing work, we do not assume that there is only one vulnerable product on each host, and instead we adopted a multi-label model to represent various attack vectors on each host, offered by different products.  Such a model is of great help to investigate the collaboration of multiple exploits. %\tlnote{revised this paragraph} %As demonstrated in the Stuxnet case, when equipped with multiple zero-day exploits, attackers can spread the malware over a network more wider and quicker.

We formally represented the network by a MRF model with different services (encoded as labels) and products (encoded as values) for each host. Such a model can then be efficiently optimized by the TRW-S algorithm. Thus, we can obtain an optimal assignment of products for a given network. The optimal solution is able to maximize the defense strength of the network against malware propagation. Compared to random diversification plans, the optimal solution is more effective in cutting off valid attack paths. In the scalability analysis, we illustrated that our method scaled well in large-scale high-density networks.

We contend that our approach has great value and potential in practical applications, by which we can advise on the best diversification strategy for a system operator to decide the most robust way to upgrade an existing ICS. We also demonstrated the practical usage of our optimization approach in a realistic case study. Furthermore, we provided a way to specify configuration constraints that we might encounter in practice. Constrained optimal solutions can be produced to accommodate these constraints.
%The optimization converges within 3 seconds for up to 1000 hosts in a network.

There are several promising lines of research to carry on. The vulnerability similarity of products in this work is estimated by data from CVE/NVD database. We are aware of the potential ``publication bias'' of CVE/NVD. However, as discussed in  \cite{johnson2016can}, NVD is currently the most trustworthy database, compared to the others. One of our key contributions is the introduction of the similarity metric and the actual use of the metric in our optimization in a way that we can more accurately capture the spread of zero-day malware. We will keep working on finding more convincing sources to provide values for this metric. Besides, we are also working on a more systematic way to estimate the vulnerability similarity between two products, such as
\begin{inparaenum}[(i)]
\item from the perspective of software engineering to analyze difference of the exploits for different products \cite{calleja2016look}; or
\item by estimating how diverse two products are \cite{nayak2014some}.
\end{inparaenum}
Besides, as our approach provides highly competitive efficiency and scalability, we are looking at optimal diversification for dynamic networks.
%\tlnote{conflicts, contraditionary between constrains, or with policies}

\bibliographystyle{ACM-Reference-Format}
\bibliography{tdsc17}


\begin{thebibliography}{32}

%%% the Web address.

\ifx \showCODEN    \undefined \def \showCODEN     #1{\unskip}     \fi
\ifx \showDOI      \undefined \def \showDOI       #1{#1}\fi
\ifx \showISBNx    \undefined \def \showISBNx     #1{\unskip}     \fi
\ifx \showISBNxiii \undefined \def \showISBNxiii  #1{\unskip}     \fi
\ifx \showISSN     \undefined \def \showISSN      #1{\unskip}     \fi
\ifx \showLCCN     \undefined \def \showLCCN      #1{\unskip}     \fi
\ifx \shownote     \undefined \def \shownote      #1{#1}          \fi
\ifx \showarticletitle \undefined \def \showarticletitle #1{#1}   \fi
\ifx \showURL      \undefined \def \showURL       {\relax}        \fi
% invisible to TeX
\providecommand\bibfield[2]{#2}
\providecommand\bibinfo[2]{#2}
\providecommand\natexlab[1]{#1}
\providecommand\showeprint[2][]{arXiv:#2}

\bibitem[\protect\citeauthoryear{Avizienis}{Avizienis}{1985}]%
        {avizienis1985n}
\bibfield{author}{\bibinfo{person}{Algirdas Avizienis}.}
  \bibinfo{year}{1985}\natexlab{}.
\newblock \showarticletitle{The N-version approach to fault-tolerant software}.
\newblock \bibinfo{journal}{\emph{IEEE Transactions on software engineering}}
  \bibinfo{number}{12} (\bibinfo{year}{1985}), \bibinfo{pages}{1491--1501}.
\newblock


\bibitem[\protect\citeauthoryear{Baudry and Monperrus}{Baudry and
  Monperrus}{2015}]%
        {baudry2015multiple}
\bibfield{author}{\bibinfo{person}{Benoit Baudry} {and} \bibinfo{person}{Martin
  Monperrus}.} \bibinfo{year}{2015}\natexlab{}.
\newblock \showarticletitle{The multiple facets of software diversity: Recent
  developments in year 2000 and beyond}.
\newblock \bibinfo{journal}{\emph{ACM Computing Surveys (CSUR)}}
  \bibinfo{volume}{48}, \bibinfo{number}{1} (\bibinfo{year}{2015}),
  \bibinfo{pages}{16}.
\newblock


\bibitem[\protect\citeauthoryear{Bhatkar, DuVarney, and Sekar}{Bhatkar
  et~al\mbox{.}}{2003}]%
        {bhatkar2003address}
\bibfield{author}{\bibinfo{person}{Sandeep Bhatkar}, \bibinfo{person}{Daniel~C
  DuVarney}, {and} \bibinfo{person}{Ron Sekar}.}
  \bibinfo{year}{2003}\natexlab{}.
\newblock \showarticletitle{Address Obfuscation: An Efficient Approach to
  Combat a Broad Range of Memory Error Exploits.}. In
  \bibinfo{booktitle}{\emph{USENIX Security Symposium}},
  Vol.~\bibinfo{volume}{12}. \bibinfo{pages}{291--301}.
\newblock


\bibitem[\protect\citeauthoryear{Borbor, Wang, Jajodia, and Singhal}{Borbor
  et~al\mbox{.}}{2016}]%
        {borbor2016diversifying}
\bibfield{author}{\bibinfo{person}{Daniel Borbor}, \bibinfo{person}{Lingyu
  Wang}, \bibinfo{person}{Sushil Jajodia}, {and} \bibinfo{person}{Anoop
  Singhal}.} \bibinfo{year}{2016}\natexlab{}.
\newblock \showarticletitle{Diversifying Network Services Under Cost
  Constraints for Better Resilience Against Unknown Attacks}. In
  \bibinfo{booktitle}{\emph{IFIP Annual Conference on Data and Applications
  Security and Privacy}}. Springer, \bibinfo{pages}{295--312}.
\newblock


\bibitem[\protect\citeauthoryear{Bozorgi, Saul, Savage, and Voelker}{Bozorgi
  et~al\mbox{.}}{2010}]%
        {bozorgi2010beyond}
\bibfield{author}{\bibinfo{person}{Mehran Bozorgi}, \bibinfo{person}{Lawrence~K
  Saul}, \bibinfo{person}{Stefan Savage}, {and} \bibinfo{person}{Geoffrey~M
  Voelker}.} \bibinfo{year}{2010}\natexlab{}.
\newblock \showarticletitle{Beyond heuristics: learning to classify
  vulnerabilities and predict exploits}. In
  \bibinfo{booktitle}{\emph{Proceedings of the 16th ACM SIGKDD international
  conference on Knowledge discovery and data mining}}. ACM,
  \bibinfo{pages}{105--114}.
\newblock


\bibitem[\protect\citeauthoryear{Byres, Ginter, and Langill}{Byres
  et~al\mbox{.}}{[n.d.]}]%
        {stuxnet-prop}
\bibfield{author}{\bibinfo{person}{Eric Byres}, \bibinfo{person}{Andrew
  Ginter}, {and} \bibinfo{person}{Joel Langill}.}
  \bibinfo{year}{[n.d.]}\natexlab{}.
\newblock \bibinfo{booktitle}{\emph{How Stuxnet Spreads – A Study of
  Infection Paths in Best Practice Systems}}.
\newblock
\newblock
\shownote{Available at
  \url{https://www.tofinosecurity.com/how-stuxnet-spreads}, access date:
  February 09, 2018.}


\bibitem[\protect\citeauthoryear{Calleja, Tapiador, and Caballero}{Calleja
  et~al\mbox{.}}{2016}]%
        {calleja2016look}
\bibfield{author}{\bibinfo{person}{Alejandro Calleja}, \bibinfo{person}{Juan
  Tapiador}, {and} \bibinfo{person}{Juan Caballero}.}
  \bibinfo{year}{2016}\natexlab{}.
\newblock \showarticletitle{A Look into 30 Years of Malware Development from a
  Software Metrics Perspective}. In \bibinfo{booktitle}{\emph{International
  Symposium on Research in Attacks, Intrusions, and Defenses}}. Springer,
  \bibinfo{pages}{325--345}.
\newblock


\bibitem[\protect\citeauthoryear{Choi, Cha, and Tappert}{Choi
  et~al\mbox{.}}{2010}]%
        {choi2010survey}
\bibfield{author}{\bibinfo{person}{Seung-Seok Choi}, \bibinfo{person}{Sung-Hyuk
  Cha}, {and} \bibinfo{person}{Charles~C Tappert}.}
  \bibinfo{year}{2010}\natexlab{}.
\newblock \showarticletitle{A survey of binary similarity and distance
  measures}.
\newblock \bibinfo{journal}{\emph{Journal of Systemics, Cybernetics and
  Informatics}} \bibinfo{volume}{8}, \bibinfo{number}{1}
  (\bibinfo{year}{2010}), \bibinfo{pages}{43--48}.
\newblock


\bibitem[\protect\citeauthoryear{CVE-Details}{CVE-Details}{[n.d.]}]%
        {cve-details1}
\bibfield{author}{\bibinfo{person}{CVE-Details}.}
  \bibinfo{year}{[n.d.]}\natexlab{}.
\newblock \bibinfo{booktitle}{\emph{Top 50 Products By Total Number Of
  "Distinct" Vulnerabilities}}.
\newblock
\newblock
\shownote{Available at \url{http://www.cvedetails.com/top-50-products.php} and
  \url{http://www.cvedetails.com/top-50-versions.php}, access date: February
  09, 2018.}


\bibitem[\protect\citeauthoryear{Falliere, Murchu, and Chien}{Falliere
  et~al\mbox{.}}{2011}]%
        {stuxnet}
\bibfield{author}{\bibinfo{person}{Nicolas Falliere}, \bibinfo{person}{Liam~O
  Murchu}, {and} \bibinfo{person}{Eric Chien}.}
  \bibinfo{year}{2011}\natexlab{}.
\newblock \showarticletitle{W32. stuxnet dossier}.
\newblock \bibinfo{journal}{\emph{White paper, Symantec Corp., Security
  Response}}  \bibinfo{volume}{5} (\bibinfo{year}{2011}).
\newblock


\bibitem[\protect\citeauthoryear{Garcia, Bessani, Gashi, Neves, and
  Obelheiro}{Garcia et~al\mbox{.}}{2011}]%
        {garcia2011diversity}
\bibfield{author}{\bibinfo{person}{Miguel Garcia}, \bibinfo{person}{Alysson
  Bessani}, \bibinfo{person}{Ilir Gashi}, \bibinfo{person}{Nuno Neves}, {and}
  \bibinfo{person}{Rafael Obelheiro}.} \bibinfo{year}{2011}\natexlab{}.
\newblock \showarticletitle{OS diversity for intrusion tolerance: Myth or
  reality?}. In \bibinfo{booktitle}{\emph{Dependable Systems \& Networks (DSN),
  2011 IEEE/IFIP 41st International Conference on}}. IEEE,
  \bibinfo{pages}{383--394}.
\newblock


\bibitem[\protect\citeauthoryear{Geman and Geman}{Geman and Geman}{1984}]%
        {geman1984stochastic}
\bibfield{author}{\bibinfo{person}{Stuart Geman} {and} \bibinfo{person}{Donald
  Geman}.} \bibinfo{year}{1984}\natexlab{}.
\newblock \showarticletitle{Stochastic relaxation, Gibbs distributions, and the
  Bayesian restoration of images}.
\newblock \bibinfo{journal}{\emph{IEEE Transactions on pattern analysis and
  machine intelligence}} \bibinfo{number}{6} (\bibinfo{year}{1984}),
  \bibinfo{pages}{721--741}.
\newblock


\bibitem[\protect\citeauthoryear{Hole}{Hole}{2015}]%
        {hole2015diversity}
\bibfield{author}{\bibinfo{person}{Kjell~J{\o}rgen Hole}.}
  \bibinfo{year}{2015}\natexlab{}.
\newblock \showarticletitle{Diversity reduces the impact of malware}.
\newblock \bibinfo{journal}{\emph{IEEE Security \& Privacy}}
  \bibinfo{volume}{13}, \bibinfo{number}{3} (\bibinfo{year}{2015}),
  \bibinfo{pages}{48--54}.
\newblock


\bibitem[\protect\citeauthoryear{Johnson, Lagerstrom, Ekstedt, and
  Franke}{Johnson et~al\mbox{.}}{2016}]%
        {johnson2016can}
\bibfield{author}{\bibinfo{person}{Pontus Johnson}, \bibinfo{person}{Robert
  Lagerstrom}, \bibinfo{person}{Mathias Ekstedt}, {and} \bibinfo{person}{Ulrik
  Franke}.} \bibinfo{year}{2016}\natexlab{}.
\newblock \showarticletitle{Can the Common Vulnerability Scoring System be
  Trusted? A Bayesian Analysis}.
\newblock \bibinfo{journal}{\emph{IEEE Transactions on Dependable and Secure
  Computing}} (\bibinfo{year}{2016}).
\newblock


\bibitem[\protect\citeauthoryear{Kolmogorov}{Kolmogorov}{2015}]%
        {kolmogorov2015new}
\bibfield{author}{\bibinfo{person}{Vladimir Kolmogorov}.}
  \bibinfo{year}{2015}\natexlab{}.
\newblock \showarticletitle{A new look at reweighted message passing}.
\newblock \bibinfo{journal}{\emph{IEEE transactions on pattern analysis and
  machine intelligence}} \bibinfo{volume}{37}, \bibinfo{number}{5}
  (\bibinfo{year}{2015}), \bibinfo{pages}{919--930}.
\newblock


\bibitem[\protect\citeauthoryear{Larsen, Homescu, Brunthaler, and Franz}{Larsen
  et~al\mbox{.}}{2014}]%
        {larsen2014sok}
\bibfield{author}{\bibinfo{person}{Per Larsen}, \bibinfo{person}{Andrei
  Homescu}, \bibinfo{person}{Stefan Brunthaler}, {and} \bibinfo{person}{Michael
  Franz}.} \bibinfo{year}{2014}\natexlab{}.
\newblock \showarticletitle{SoK: Automated software diversity}. In
  \bibinfo{booktitle}{\emph{Security and Privacy (SP), 2014 IEEE Symposium
  on}}. IEEE, \bibinfo{pages}{276--291}.
\newblock


\bibitem[\protect\citeauthoryear{Li and Hankin}{Li and Hankin}{2016}]%
        {li2016critis}
\bibfield{author}{\bibinfo{person}{Tingting Li} {and} \bibinfo{person}{Chris
  Hankin}.} \bibinfo{year}{2016}\natexlab{}.
\newblock \showarticletitle{Effective Defence Against Zero-Day Exploits Using
  Bayesian Networks}. In \bibinfo{booktitle}{\emph{International Conference on
  Critical Information Infrastructures Security}}. Springer.
\newblock


\bibitem[\protect\citeauthoryear{MITRE}{MITRE}{[n.d.]}]%
        {cve}
\bibfield{author}{\bibinfo{person}{MITRE}.} \bibinfo{year}{[n.d.]}\natexlab{}.
\newblock \bibinfo{booktitle}{\emph{Common vulnerabilities and exposures}}.
\newblock
\newblock
\shownote{Available at \url{https://cve.mitre.org/}, last acceessed on February
  09, 2018.}


\bibitem[\protect\citeauthoryear{Moreels and Dulaunoy}{Moreels and
  Dulaunoy}{[n.d.]}]%
        {cve-search}
\bibfield{author}{\bibinfo{person}{Pieter-Jan Moreels} {and}
  \bibinfo{person}{Alexandre Dulaunoy}.} \bibinfo{year}{[n.d.]}\natexlab{}.
\newblock \bibinfo{booktitle}{\emph{\textsc{cve-search}}}.
\newblock
\newblock
\shownote{GitHub repository at \url{https://github.com/cve-search/cve-search},
  access date: February 09, 2018.}


\bibitem[\protect\citeauthoryear{Nayak, Marino, Efstathopoulos, and
  Dumitra{\c{s}}}{Nayak et~al\mbox{.}}{2014}]%
        {nayak2014some}
\bibfield{author}{\bibinfo{person}{Kartik Nayak}, \bibinfo{person}{Daniel
  Marino}, \bibinfo{person}{Petros Efstathopoulos}, {and}
  \bibinfo{person}{Tudor Dumitra{\c{s}}}.} \bibinfo{year}{2014}\natexlab{}.
\newblock \showarticletitle{Some vulnerabilities are different than others}. In
  \bibinfo{booktitle}{\emph{International Workshop on Recent Advances in
  Intrusion Detection}}. Springer, \bibinfo{pages}{426--446}.
\newblock


\bibitem[\protect\citeauthoryear{Newell, Obenshain, Tantillo, Nita-Rotaru, and
  Amir}{Newell et~al\mbox{.}}{2015}]%
        {newell2015increasing}
\bibfield{author}{\bibinfo{person}{Andrew Newell}, \bibinfo{person}{Daniel
  Obenshain}, \bibinfo{person}{Thomas Tantillo}, \bibinfo{person}{Cristina
  Nita-Rotaru}, {and} \bibinfo{person}{Yair Amir}.}
  \bibinfo{year}{2015}\natexlab{}.
\newblock \showarticletitle{Increasing network resiliency by optimally
  assigning diverse variants to routing nodes}.
\newblock \bibinfo{journal}{\emph{IEEE Transactions on Dependable and Secure
  Computing}} \bibinfo{volume}{12}, \bibinfo{number}{6} (\bibinfo{year}{2015}),
  \bibinfo{pages}{602--614}.
\newblock


\bibitem[\protect\citeauthoryear{NIST}{NIST}{[n.d.]}]%
        {nvd}
\bibfield{author}{\bibinfo{person}{NIST}.} \bibinfo{year}{[n.d.]}\natexlab{}.
\newblock \bibinfo{booktitle}{\emph{National Vulnerability Database}}.
\newblock
\newblock
\shownote{Available at \url{https://nvd.nist.gov/}, access date: February 09,
  2018.}


\bibitem[\protect\citeauthoryear{O'Donnell and Sethu}{O'Donnell and
  Sethu}{2004}]%
        {o2004achieving}
\bibfield{author}{\bibinfo{person}{Adam~J O'Donnell} {and}
  \bibinfo{person}{Harish Sethu}.} \bibinfo{year}{2004}\natexlab{}.
\newblock \showarticletitle{On achieving software diversity for improved
  network security using distributed coloring algorithms}. In
  \bibinfo{booktitle}{\emph{Proceedings of the 11th ACM conference on Computer
  and communications security}}. ACM, \bibinfo{pages}{121--131}.
\newblock


\bibitem[\protect\citeauthoryear{Pal, Schantz, Paulos, and Benyo}{Pal
  et~al\mbox{.}}{2014}]%
        {pal2014managed}
\bibfield{author}{\bibinfo{person}{Partha Pal}, \bibinfo{person}{Richard
  Schantz}, \bibinfo{person}{Aaron Paulos}, {and} \bibinfo{person}{Brett
  Benyo}.} \bibinfo{year}{2014}\natexlab{}.
\newblock \showarticletitle{Managed execution environment as a moving-target
  defense infrastructure}.
\newblock \bibinfo{journal}{\emph{IEEE Security \& Privacy}}
  \bibinfo{volume}{12}, \bibinfo{number}{2} (\bibinfo{year}{2014}),
  \bibinfo{pages}{51--59}.
\newblock


\bibitem[\protect\citeauthoryear{Pappas, Polychronakis, and Keromytis}{Pappas
  et~al\mbox{.}}{2012}]%
        {pappas2012smashing}
\bibfield{author}{\bibinfo{person}{Vasilis Pappas}, \bibinfo{person}{Michalis
  Polychronakis}, {and} \bibinfo{person}{Angelos~D Keromytis}.}
  \bibinfo{year}{2012}\natexlab{}.
\newblock \showarticletitle{Smashing the gadgets: Hindering return-oriented
  programming using in-place code randomization}. In
  \bibinfo{booktitle}{\emph{Security and Privacy (SP), 2012 IEEE Symposium
  on}}. IEEE, \bibinfo{pages}{601--615}.
\newblock


\bibitem[\protect\citeauthoryear{Pearl}{Pearl}{2014}]%
        {pearl2014probabilistic}
\bibfield{author}{\bibinfo{person}{Judea Pearl}.}
  \bibinfo{year}{2014}\natexlab{}.
\newblock \bibinfo{booktitle}{\emph{Probabilistic reasoning in intelligent
  systems: networks of plausible inference}}.
\newblock \bibinfo{publisher}{Morgan Kaufmann}.
\newblock


\bibitem[\protect\citeauthoryear{SIEMENS}{SIEMENS}{[n.d.]}]%
        {wincc}
\bibfield{author}{\bibinfo{person}{SIEMENS}.}
  \bibinfo{year}{[n.d.]}\natexlab{}.
\newblock \bibinfo{booktitle}{\emph{WinCC v7.4: General information and
  installation}}.
\newblock
\newblock
\shownote{Available at
  \url{https://cache.industry.siemens.com/dl/files/216/109736216/att_879785/v1/WinCC_GeneralInfo_Installation_Readme_en-US_en-US.pdf},
  access date: February 09, 2018.}


\bibitem[\protect\citeauthoryear{Stouffer, Pillitteri, Lightman, Abrams, and
  Hahn}{Stouffer et~al\mbox{.}}{2015}]%
        {nistguide}
\bibfield{author}{\bibinfo{person}{Keith Stouffer}, \bibinfo{person}{Victoria
  Pillitteri}, \bibinfo{person}{Suzanne Lightman}, \bibinfo{person}{Marshall
  Abrams}, {and} \bibinfo{person}{Adam Hahn}.} \bibinfo{year}{2015}\natexlab{}.
\newblock \showarticletitle{Guide to Industrial Control Systems (ICS)
  Security}.
\newblock \bibinfo{journal}{\emph{NIST Special Publication}}
  \bibinfo{volume}{800} (\bibinfo{year}{2015}), \bibinfo{pages}{82}.
\newblock


\bibitem[\protect\citeauthoryear{Wang, Jajodia, Singhal, and Noel}{Wang
  et~al\mbox{.}}{2010}]%
        {wang2010k}
\bibfield{author}{\bibinfo{person}{Lingyu Wang}, \bibinfo{person}{Sushil
  Jajodia}, \bibinfo{person}{Anoop Singhal}, {and} \bibinfo{person}{Steven
  Noel}.} \bibinfo{year}{2010}\natexlab{}.
\newblock \showarticletitle{k-zero day safety: Measuring the security risk of
  networks against unknown attacks}.
\newblock In \bibinfo{booktitle}{\emph{Computer Security--ESORICS 2010}}.
  \bibinfo{publisher}{Springer}, \bibinfo{pages}{573--587}.
\newblock


\bibitem[\protect\citeauthoryear{Wang, Zhang, Jajodia, Singhal, and
  Albanese}{Wang et~al\mbox{.}}{2014}]%
        {wang2014modeling}
\bibfield{author}{\bibinfo{person}{Lingyu Wang}, \bibinfo{person}{Mengyuan
  Zhang}, \bibinfo{person}{Sushil Jajodia}, \bibinfo{person}{Anoop Singhal},
  {and} \bibinfo{person}{Massimiliano Albanese}.}
  \bibinfo{year}{2014}\natexlab{}.
\newblock \showarticletitle{Modeling Network Diversity for Evaluating the
  Robustness of Networks against Zero-Day Attacks}.
\newblock In \bibinfo{booktitle}{\emph{Computer Security-ESORICS 2014}}.
  \bibinfo{publisher}{Springer}, \bibinfo{pages}{494--511}.
\newblock


\bibitem[\protect\citeauthoryear{Wilensky}{Wilensky}{[n.d.]}]%
        {netlogo}
\bibfield{author}{\bibinfo{person}{U. Wilensky}.}
  \bibinfo{year}{[n.d.]}\natexlab{}.
\newblock \bibinfo{booktitle}{\emph{NetLogo}}.
\newblock
\newblock
\shownote{Available at \url{http://ccl.northwestern.edu/netlogo/.}, access
  date: February 09, 2018.}


\bibitem[\protect\citeauthoryear{Zhang, Wang, Jajodia, Singhal, and
  Albanese}{Zhang et~al\mbox{.}}{2016}]%
        {zhang2016network}
\bibfield{author}{\bibinfo{person}{Mengyuan Zhang}, \bibinfo{person}{Lingyu
  Wang}, \bibinfo{person}{Sushil Jajodia}, \bibinfo{person}{Anoop Singhal},
  {and} \bibinfo{person}{Massimiliano Albanese}.}
  \bibinfo{year}{2016}\natexlab{}.
\newblock \showarticletitle{Network diversity: a security metric for evaluating
  the resilience of networks against zero-day attacks}.
\newblock \bibinfo{journal}{\emph{IEEE Transactions on Information Forensics
  and Security}} \bibinfo{volume}{11}, \bibinfo{number}{5}
  (\bibinfo{year}{2016}), \bibinfo{pages}{1071--1086}.
\newblock


\end{thebibliography}

\end{document}